\documentclass[a4paper,twoside,english,aps,preprint,showpacs]{revtex4}
\usepackage{times}
\usepackage[T1]{fontenc}
\usepackage[latin1]{inputenc}
\usepackage{subfigure}
\usepackage{amsmath}
\usepackage{graphicx}
\usepackage{amssymb}

\makeatletter

\newcommand{\noun}[1]{\textsc{#1}}

\usepackage{babel}
\makeatother
\begin{document}

\title{Correlations in Ultracold Trapped Few-Boson Systems: Transition from
Condensation to Fermionization}

\author{Sascha Zöllner}

\email{sascha.zoellner@pci.uni-heidelberg.de}

\affiliation{Theoretische Chemie, Institut f\"{u}r Physikalische Chemie, Universit\"{a}t
Heidelberg, INF 229, 69120 Heidelberg, Germany }

\author{Hans-Dieter Meyer}

\email{hans-dieter.meyer@pci.uni-heidelberg.de}

\affiliation{Theoretische Chemie, Institut f\"{u}r Physikalische Chemie, Universit\"{a}t
Heidelberg, INF 229, 69120 Heidelberg, Germany}

\author{Peter Schmelcher}

\email{peter.schmelcher@pci.uni-heidelberg.de}

\affiliation{Theoretische Chemie, Institut f\"{u}r Physikalische Chemie, Universit\"{a}t
Heidelberg, INF 229, 69120 Heidelberg, Germany}

\affiliation{Physikalisches Institut, Universit\"{a}t Heidelberg, Philosophenweg
12, 69120 Heidelberg, Germany}


\begin{abstract}
We study the correlation properties of the ground states of few ultracold
bosons, trapped in double wells of varying barrier height in one dimension.
Extending previous results on the signature of the transition from
a Bose-condensed state via fragmentation to the hard-core limit, we
provide a deeper understanding of that transition by relating it to
the loss of coherence in the one-body density matrix and to the emerging
long-range tail in the momentum spectrum. These are accounted for
in detail by discussing the natural orbitals and their occupations.
Our discussion is complemented by an analysis of the two-body correlation
function.
\end{abstract}
\maketitle

\section{Introduction}

For more than a decade, trapped ultracold atoms have been a vastly
expanding field of research \cite{pitaevskii,dalfovo99,pethick,leggett01}.
Their low thermal fluctuations, and the fact that one can virtually
design both the external as well as the inter-particle forces via
electromagnetic fields, make them an ideal simulation tool for various
phenomena, ranging from condensed-matter physics to quantum information.
A particularly intriguing aspect has been the system's dimensionality.
For example, if its transverse degrees of freedom are frozen out such
that an effective one-dimensional description becomes possible, then
the effective interaction strength of the system may be tuned at will
from a weakly correlated to a strongly interacting one by merely changing
the lengthscale of the transverse confinement \cite{Olshanii1998a}.
In the case of infinite repulsion---the so-called Tonks-Girardeau
gas---the system may even be mapped to that of an ideal Fermi gas
\cite{girardeau60} and exhibits striking features, including the
reduction of off-diagonal long-range order \cite{vaidya79} or a very
distinctive momentum spectrum \cite{minguzzi02}. This has also been
verified experimentally \cite{kinoshita04,paredes04}.

However, the Tonks-Girardeau limit requires low densities and is thus
amenable only for \emph{few} atoms, typically $N\sim10-100$. Besides
that, there is the factor of computational accessibility---which includes
the fact that a small number of atoms greatly facilitates the intuitive
understanding of the mechanisms underlying the physics of \emph{large}
systems. But there is yet another argument for considering few-atom
systems: They permit a much higher level of control. There is no thermal
cloud, as for large $N$, associated with decoherence and energetically
dense excitations, but a pure quantum system.

However, the same reasons that make few atoms so interesting also
render their solution computationally cumbersome. There have been
some detailed analyses of the Tonks-Girardeau limit in a harmonic
trap \cite{girardeau01,papenbrock03}, which is greatly simplified
since the solution is semi-analytic. Needless to say, the complementary
borderline case of weakly interacting bosons is also well understood:
in this limit, the Gross-Pitaevskii equation becomes valid, which
assumes that all particles condense into a single one-particle orbital
(see, e.g., \cite{pitaevskii}). Comparatively little is known about
the \emph{transition} between these extremes, though. Many key features
are nicely illustrated on the analytic solution of two bosons in a
harmonic trap \cite{cirone01} or other simple models \cite{hao06,sakmann05}.
Based on a multi-orbital mean-field approach valid for arbitrary traps,
it has recently been demonstrated that there exists an intriguing
pathway for the above transition \cite{alon05}. While that ansatz
captures the essential features of that evolution, it is desirable
to investigate it from a rigorous many-body perspective, but still
irrespective of the trap geometry. Most of the studies so far have
relied on the assumption of only few contributing orbitals \cite{masiello05,streltsov06},
while a recent exact diagonalization has focused on a simple harmonic
trap \cite{deuretzbacher06}. For the case of a double-well trap,
the authors have previously investigated the ground state of few interacting
bosons in the entire regime between the Gross-Pitaevskii limit and
fermionization for different well depths, but chiefly from the perspective
of the density profiles \cite{zoellner06a}.

The aim of this paper now is to address the role of correlations in
these systems, in particular the relation of the fragmentation mechanism
to the concept of long-range order and the momentum density. This
way we not only bridge the gap between the two borderline regimes
of zero and infinite repulsion but also extend our previous study.
Our approach is based on the Multi-Configuration Time-Dependent Hartree
method \cite{mey03:251,mey98:3011,bec00:1}, which we use to study
the numerically exact ground state of few bosons.

This article is organized as follows. In Sec.~\ref{sec:theory},
the model is introduced and some key quantities for the description
of correlations are discussed. Section~\ref{sec:method} contains
a concise introduction to the computational method and how it can
be applied to our problem. In the subsequent section the one-body
correlation aspects are studied. More concretely, we present the full
one-body density matrix in the context of long-range order in Sec.~\ref{sub:ODLRO},
which is subsequently explained in terms of its eigenvectors and their
populations (\ref{sub:natorb}). Their connection to the momentum
density will then be clarified in Sec.~\ref{sub:Momentum-density}.
We complement our investigation by going beyond the one-particle picture
in looking at the two-body correlations in Sec.~\ref{sec:2p-correlation},
which is rounded off by relating our results to some common approximation
schemes, the two-mode model \cite{spekkens99} and the multi-orbital
mean field.

\section{Theoretical background \label{sec:theory}}

\subsection{The model}

 In this article we investigate a system of \emph{few} interacting
bosons ($N=2,\dots,6$) in an external trap. These particles, representing
atoms with mass $M$, are taken to be one-dimensional (1D). More precisely,
after integrating out the transverse degrees of freedom and rescaling
we arrive at the model Hamiltonian (see \cite{zoellner06a} for details)
\[
H=\sum_{i}h_{i}+\sum_{i<j}V(x_{i}-x_{j}),\]
where $h=\frac{1}{2}p^{2}+U(x)$ is the one-particle Hamiltonian with
a trapping potential $U$, while $V$ is the two-particle interaction
potential with low-energy scattering length $a_{0}$, taken to be
the effective interaction \cite{Olshanii1998a}\[
V(x)=g\delta_{\sigma}(x),\textrm{ with }g=\frac{4a_{0}}{a_{\perp}^{2}}\left(1-\left|\zeta\left({\scriptstyle \frac{1}{2}}\right)\right|\frac{a_{0}}{a_{\perp}}\right)^{-1}\negthickspace.\]
Here a harmonic transverse trap potential with oscillator length $a_{\perp}$
was assumed. Moreover, the well-known numerical difficulties due to
the spurious short-range behavior of the delta-function potential
$\delta(x)$ are alleviated by mollifying it with the normalized Gaussian\[
\delta_{\sigma}(x)=\frac{1}{\sqrt{2\pi}\sigma}e^{-x^{2}/2\sigma^{2}},\]
which tends to $\delta$ as $\sigma\to0$ in the distribution sense.
We choose a fixed value $\sigma=.05$ as a trade-off between smoothness
and a short range.

\subsection{Fragmentation: key aspects\label{sub:Fragmentation}}

Although our approach equips us with the full solution of the system---here,
the ground-state wave function---this solution obviously still needs
to be related to the concrete physical questions. Penrose and Onsager
suggested a criterion connected with the one-body density matrix,
which will be laid in what follows.

As is well-known, the knowledge of the wave function $\Psi$ is equivalent
to that of the density matrix $\rho_{N}=|\Psi\rangle\langle\Psi|$.
To the extent that we study at most two-body correlations, it already
suffices to consider the two-particle density operator\begin{equation}
\rho_{2}=\mathrm{tr}_{3..N}|\Psi\rangle\langle\Psi|,\label{eq:rho2}\end{equation}
whose diagonal kernel $\rho_{2}(x_{1},x_{2})$ gives the probability
density for finding one particle located at $x_{1}$ and any second
one at $x_{2}$. For any 1-particle operator, of course, it would
be enough to know the one-particle density matrix $\rho_{1}=\mathrm{tr}_{2}\rho_{2}$,
so that the exact ground-state energy may be written as\[
E=N\,\mathrm{tr}(\rho_{1}h)+\frac{N(N-1)}{2}\mathrm{tr}(\rho_{2}V).\]

Consider the spectral decomposition of the one-particle density matrix
\begin{equation}
\rho_{1}\equiv\sum_{a}n_{a}|\phi_{a}\rangle\langle\phi_{a}|,\label{eq:rho1}\end{equation}
 where $n_{a}\in[0,1]$ is said to be the population of the \textit{natural
orbital} $\phi_{a}$. If all $n'_{a}\equiv n_{a}N\in\mathbb{N}$ ($\sum_{a}n_{a}'=N$),
then the density may be mapped to the (non-interacting) number state
$|n'_{0},n'_{1},\dots\rangle$ based on the one-particle basis $\{\phi_{a}\}$;
for non-integer values it extends that concept. In particular, the
highest such occupation, $n_{0}$, may serve as a measure of \emph{non}-fragmentation,
a criterion put forward by Penrose and Onsager \cite{Penrose56}.
For $n_{0}=1$, a simple condensate is recovered. This is the well-known
borderline case of the Gross-Pitaevskii eq.: as $g\rightarrow0$,
$\rho_{1}\rightarrow|\phi_{0}\rangle\langle\phi_{0}|$ \cite{lieb03}
and $\rho_{2}=\rho_{1}\otimes\rho_{1}$, so that the interaction above
can be replaced by a mean field $\bar{V}=\mathrm{tr}(\rho_{1}V)$. 

The 1-particle density matrix may be viewed not only from the perspective
of its spectral decomposition, but also in terms of its integral kernel
$\rho_{1}(x,x')\equiv\langle x|\rho_{1}|x'\rangle=\rho_{1}(x',x)^{*}$.
Since the density matrix is non-negative, so is the 1-particle density
$\rho(x)\equiv\rho_{1}(x,x)$. As opposed to that, the off-diagonal
part will be complex in general (though in this paper a real representation
is employed). It is therefore certainly not an observable in its own
right. Nonetheless, it is of some interest as it gives us access to
all one-particle quantities, also non-local ones such as the momentum
density $\tilde{\rho}(k)=2\pi\langle k|\rho_{1}|k\rangle=\sum_{a}n_{a}|\hat{\phi}_{a}(k)|^{2}$,
which can be related to the density matrix via\[
\tilde{\rho}(k)=\int dx\int dx'e^{-ik(x-x')}\rho_{1}(x,x').\]
It is reflection symmetric if $\rho_{1}$ is real symmetric. Moreover,
it can be understood as the Fourier transform of the integrated {}`off-diagonal'
correlation function \cite{pitaevskii}\[
\tilde{\rho}(k)=\int dre^{-ikr}\gamma(r),\]
with $\gamma(r):=\int dR\rho_{1}(R+\frac{r}{2},R-\frac{r}{2})$. Note
that $\gamma$ is again generally complex and reflection symmetric,
while $\gamma(0)=\mathrm{1}$. From this, it becomes clear that the
off-diagonal behavior encoded in $\gamma$ has a 1-1 correspondence
to the momentum distribution. More specifically, the short-distance
behavior determines the high-$k$ asymptotics, which for a delta-type
interaction $V(x)=g\delta(x)$ in the limit $g\to\infty$ has been
shown to display the universal decay $\tilde{\rho}(k)=O(k^{-4})$
\cite{minguzzi02}. Conversely, the off-diagonal asymptotics $r\to\infty$
relates to the low-$k$ regime. This, however, depends on the nature
of the external potential. For a translationally invariant system,
it has been argued that Bose condensation were equivalent to \emph{off-diagonal
long-range order}, i.e. $\gamma(r)=O(1)$ \cite{yang62}. In the same
context, but in the limit $g\to\infty$, it has in turn been shown
that $\gamma(r)=O(r^{-1/2})$, which implies an infrared divergence
$\tilde{\rho}(k)\sim c/\sqrt{k}$ as $k\to0$ \cite{vaidya79}.

The above limit $g\to\infty$ is commonly referred to as the \emph{Tonks-Girardeau}
limit of 1D hard-core bosons, or also as their \emph{fermionization}.
This lingo finds its justification in the \emph{Bose-Fermi map} \cite{girardeau60,yukalov05}
that establishes an isomorphy between the exact \emph{bosonic} wave
function $\Psi_{\infty}^{+}$ and that of a (spin-polarized) non-interacting
\emph{fermionic} solution $\Psi_{0}^{-}$, \[
\Psi_{\infty}^{+}=A\Psi_{0}^{-},\]
where $A(Q)=\prod_{i<j}\mathrm{sgn}(x_{i}-x_{j})$ and $Q\equiv(x_{1},\dots,x_{N})^{T}$.
In particular, the ground state reduces simply to the absolute value
of $\Psi_{0}^{-}$, which makes it tempting to think of the hard-core
interaction $g\to\infty$ as mimicking the exclusion principle. As
the free fermionic solution is easily accessible, this theorem has
proven very fruitful in a wide range of applications. For our purposes,
the most relevant one is the solution of $N$ bosons in a harmonic
trap \cite{girardeau01}\[
\Psi_{\infty}^{+}(Q)\propto e^{-|Q|^{2}/2}\negthickspace\prod_{1\le i<j\le N}\negthickspace|x_{i}-x_{j}|,\]
which illustrates the characteristic short-distance correlations.

\section{Computational method\label{sec:method}}

Our goal is to investigate the ground state of the system introduced
in Sec.~\ref{sec:theory} for all relevant interaction strengths
in a numerically \emph{exact, i.e.,} controllable fashion. This is
a highly challenging and time-consuming task, and only few such studies
on ultracold atoms exist even for model systems (see, e.g., \cite{streltsov06,deuretzbacher06,masiello05}).
Our approach relies on the Multi-Configuration Time-Dependent Hartree
\noun{(mctdh)} method \cite{mey90:73,bec00:1,mey03:251}, primarily
a wave-packet dynamics tool known for its outstanding efficiency in
high-dimensional applications. To be self-contained, we will provide
a concise introduction to this method and how it can be adapted to
our purposes.

The underlying idea of MCTDH is to solve the time-dependent Schrödinger
equation\begin{equation}
\left\{ \begin{array}{c}
i\dot{\Psi}=H\Psi\\
\Psi(Q,0)=\Psi_{0}(Q)\end{array}\right.\label{eq:TDSE}\end{equation}
 as an initial-value problem by expansion in terms of direct (or Hartree)
products $\Phi_{J}$:\begin{equation}
\Psi(Q,t)=\sum_{J}A_{J}(t)\Phi_{J}(Q,t)\equiv\sum_{j_{1}=1}^{n_{1}}\ldots\sum_{j_{f}=1}^{n_{f}}A_{j_{1}\ldots j_{f}}(t)\prod_{\kappa=1}^{f}\varphi_{j_{\kappa}}^{(\kappa)}(x_{\kappa},t),\label{eq:mctdh-ansatz}\end{equation}
using a convenient multi-index notation for the configurations, $J=(j_{1}\dots j_{f})$,
where $f$ denotes the number of degrees of freedom and $Q\equiv(x_{1},\dots,x_{f})^{T}$.
The (unknown) \emph{single-particle functions} $\varphi_{j_{\kappa}}^{(\kappa)}$
are in turn represented in a fixed, primitive basis implemented on
a grid. For indistinguishable particles as in our case, the sets of
single-particle functions for each degree $\kappa=1,\dots,N$ are
of course identical (i.e., we have $\varphi_{j_{\kappa}}$, with $j_{\kappa}\le n$).

Note that in the above expansion, not only the coefficients $A_{J}$
are time-dependent, but so are the Hartree products $\Phi_{J}$. Using
the Dirac-Frenkel variational principle, one can derive equations
of motion for both $A_{J},\Phi_{J}$ \cite{bec00:1}. Integrating
this differential-equation system allows one to obtain the time evolution
of the system via (\ref{eq:mctdh-ansatz}). Let us emphasize that
the conceptual complication above offers an enormous advantage: the
basis $\{\Phi_{J}(\cdot,t)\}$ is variationally optimal at each time
$t$. Thus it can be kept fairly small, rendering the procedure very
efficient.

It goes without saying that the basis vectors $\Phi_{J}$ are not
permutation symmetric, as would be an obvious demand when dealing
with bosons. However, the symmetry can be enforced on $\Psi$ by symmetrizing
the coefficients $A_{J}$, even though this turns out to be unnecessary
as long as identical single-particle functions are employed.

The Heidelberg \noun{mctdh} package \cite{mctdh:package}, which
we use, incorporates a significant extension to the basic concept
outlined so far. The so-called \emph{relaxation method} \cite{kos86:223}
provides a way to not only \emph{propagate} a wave packet, but also
to obtain the lowest \emph{eigenstates} of the system. The underlying
idea is to propagate some wave function $\Psi_{0}$ by the non-unitary
$e^{-H\tau}$ (\emph{propagation in imaginary time}.) As $\tau\to\infty$,
this automatically damps out any contribution but that stemming from
the true ground state $|J=\mathbf{0}\rangle$,\[
e^{-H\tau}\Psi_{0}=\sum_{J}e^{-E_{J}\tau}|J\rangle\langle J|\Psi_{0}\rangle.\]
 In practice, one relies on a more sophisticated scheme termed \emph{improved
relaxation}. Here $\langle\Psi|H-E|\Psi\rangle$ is minimized with
respect to both the coefficients $A_{J}$ and the configurations $\Phi_{J}$.
The equations of motion thus obtained are then solved iteratively
by first solving for $A_{J}$ (by diagonalization of $(\langle\Phi_{J}|H|\Phi_{K}\rangle)$
with fixed $\Phi_{J}$) and then propagating $\Phi_{J}$ in imaginary
time over a short period. That cycle will then be repeated. The improved-relaxation
method is outlined in Ref.~\cite{mey03:251}; a more comprehensive
account is also available \cite{meyer06}.

As it stands, the effort of this method scales exponentially with
the number of degrees of freedom, $n^{N}$. This restricts our analysis
in the current setup to about $N=O(10)$, depending on how decisive
correlation effects are. If these are indeed essential, it has been
demonstrated \cite{zoellner06a} that at least $n=N$ orbitals are
needed for \emph{qualitative} convergence alone, while the true behavior
may necessitate many more. As an illustration, we consider systems
with $N\sim5$ and use $n\sim15$ orbitals. By contrast, the dependence
on the primitive basis, and thus on the grid points, is not as severe.
In our case, the grid spacing should of course be small enough to
sample the interaction potential. We consider a discrete variable
representation \cite{lig85:1400} of 95 harmonic-oscillator functions,
which is equivalent to $95$ grid points.

\section{One-body correlations \label{sec:1p-correlation}}

\begin{figure}
\begin{center}\includegraphics[%
  width=7.5cm,
  keepaspectratio]{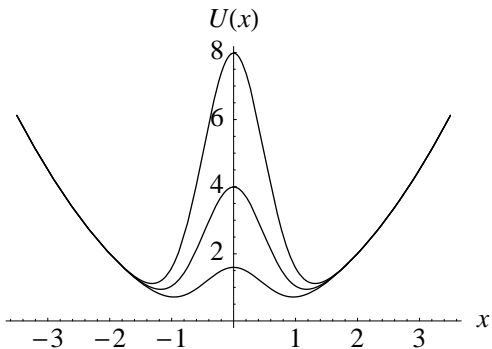}\end{center}

\caption{Sketch of the model potential $U(x)=\frac{1}{2}x^{2}+h\delta_{w}(x)$,
consisting of a harmonic trap plus a normalized Gaussian of width
$w=0.5$ and barrier strengths $h=2,5,10$. \label{cap:DWplot}}
\end{figure}

As in Ref.~\cite{zoellner06a}, we consider the ground-state properties
of bosons in a double-well trap modeled by\[
U(x)=\frac{1}{2}x^{2}+h\delta_{w}(x).\]
This potential is a superposition of a \emph{}harmonic oscillator
(HO), which it equals asymptotically, and a central barrier which
splits the trap into two fragments (Fig.~\ref{cap:DWplot}). The
barrier is shaped as a normalized Gaussian $\delta_{w}$ of width
$w=0.5$ and {}`barrier strength' $h$. 

For $h=0$, the case of interacting bosons in a harmonic trap is reproduced.
In Ref.~\cite{zoellner06a}, we have described the transition from
a simple, weakly interacting condensate ($g\to0$) to fragmentation
and finally the \emph{Tonks-Girardeau} limit ($g\rightarrow\infty$).
As $h\rightarrow\infty$, the energy barrier will greatly exceed the
energy available to the atoms, and we end up with two \textbf{\emph{}}\emph{isolated
wells}. A larger $g$ then affects only the fragmentation \emph{within}
each of these wells. In between, there is an interesting interplay
between the `static' barrier ($h$) and `dynamical barriers' in the
form of inter-particle forces ($g$), which has been analyzed mainly
from the viewpoint of spatial \emph{densities}. 

We now seek to extend that investigation to non-local properties,
such as the off-diagonal density matrix (Sec.~\ref{sub:ODLRO}) and
the momentum density (\ref{sub:Momentum-density}), so as to attain
a deeper insight into the nature of the above transition. The role
played by correlations will also be highlighted by showing how \emph{all}
the natural populations evolve as a function of $g$.

\subsection{One-particle density matrix and long-range order \label{sub:ODLRO}}

\begin{figure}
\includegraphics[%
  width=5cm,
  keepaspectratio]{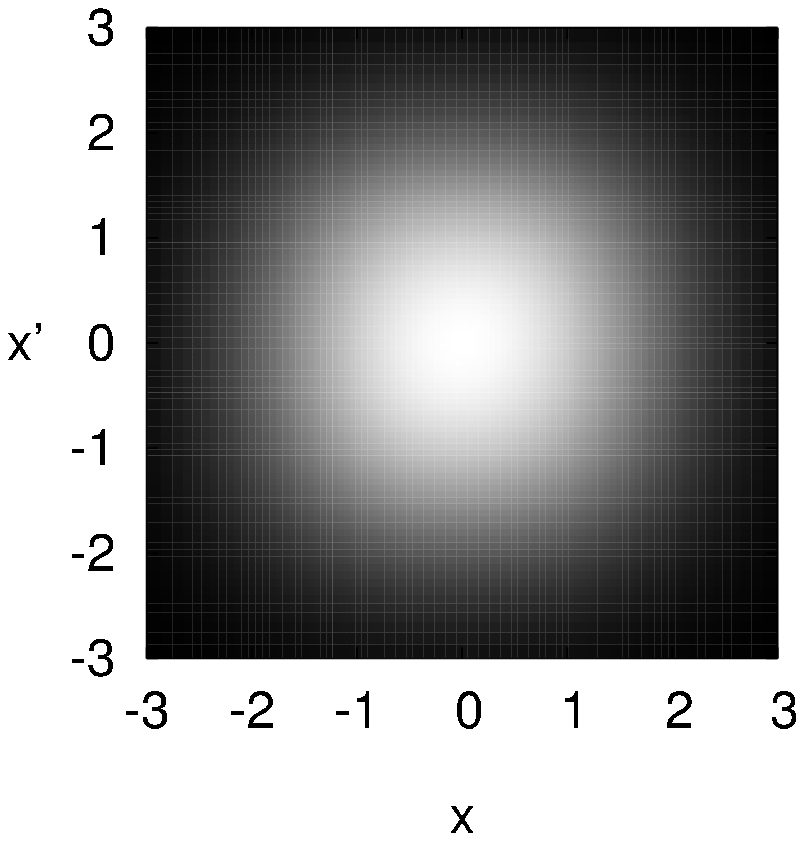}\includegraphics[%
  width=5cm,
  keepaspectratio]{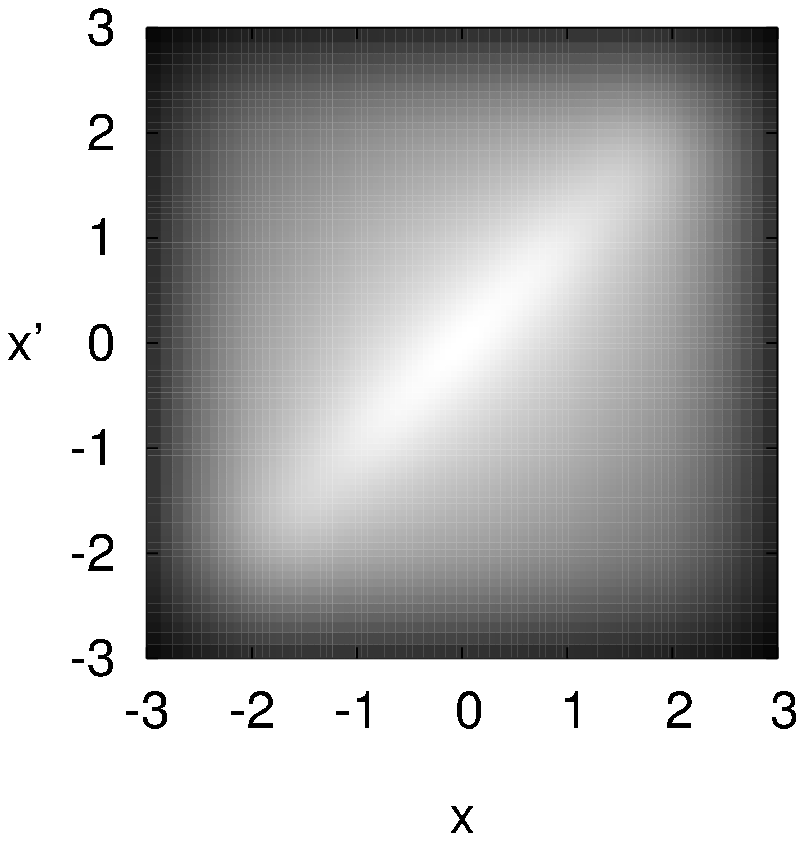}\includegraphics[%
  width=5cm,
  keepaspectratio]{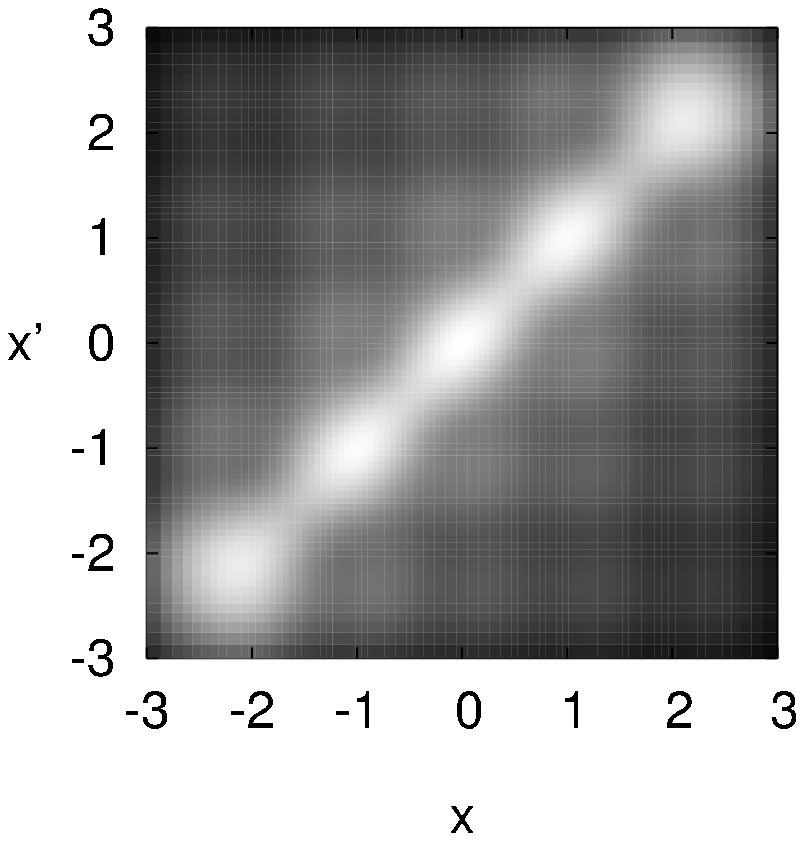}

\includegraphics[%
  width=5cm,
  keepaspectratio]{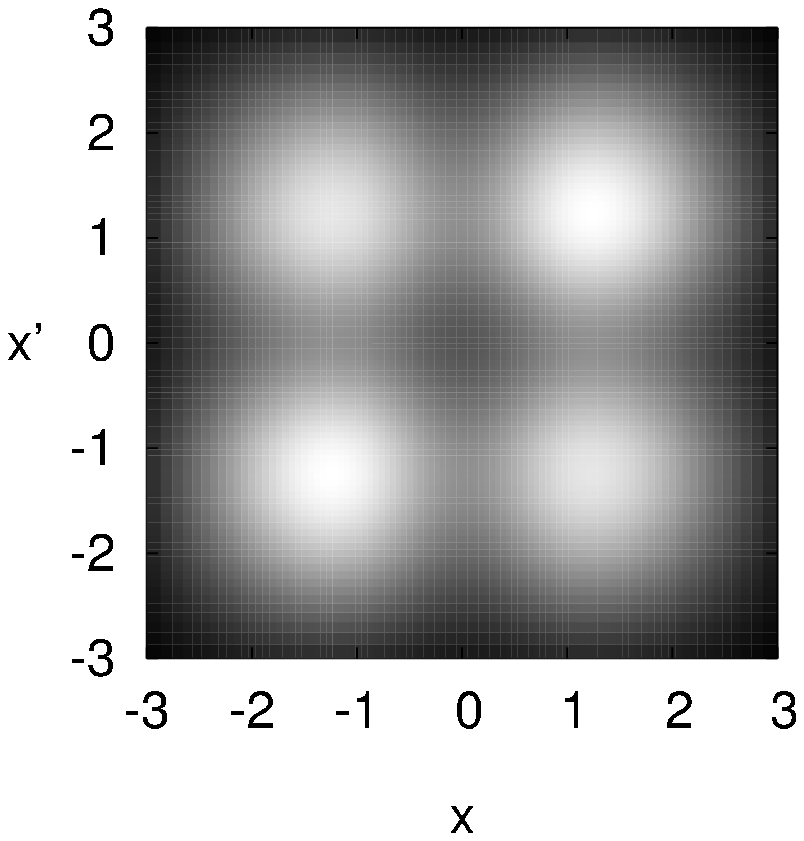}\includegraphics[%
  width=5cm,
  keepaspectratio]{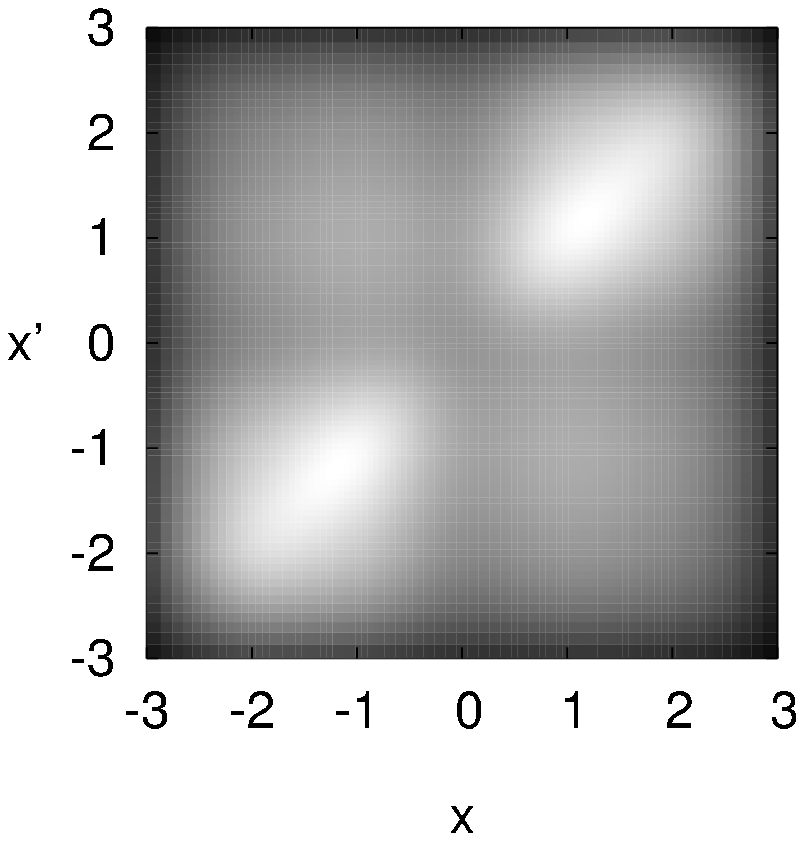}\includegraphics[%
  width=5cm,
  keepaspectratio]{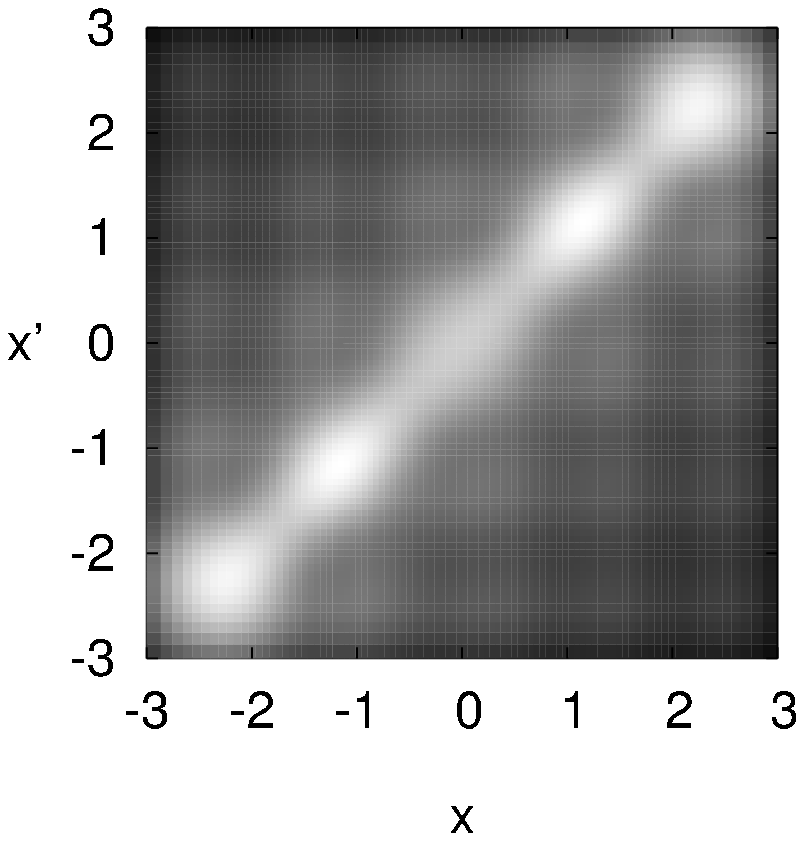}

\caption{1-particle density matrix $\rho_{1}(x,x')$ for $N=5$ bosons. Top
row: harmonic trap, bottom row: double well (barrier height $h=5$).
Results are shown for the interaction strengths $g=0.4,\,4.7,\,194$
from left to right. \label{cap:ODLRO}}
\end{figure}
The 1-particle density matrix $\rho_{1}$ contains all the information
about the one-particle aspects of the system, and serves as a good
measure for the degree of fragmentation. In this section, we will
analyze it from the most immediate perspective, i.e., we investigate
its integral kernel $\rho_{1}(x,x')$. Although it is not an observable
in itself, being generally complex-valued, it is indirectly accessible
e.g. via interferometry experiments \cite{pitaevskii}. More importantly,
the density matrix relates to salient questions such as off-diagonal
long-range order or the momentum distribution (see \ref{sub:Fragmentation}).
Insofar it serves as a good starting point for our discussion of fragmentation
in double-well systems.

In Figure~\ref{cap:ODLRO}, the fragmentation transition as reflected
in $\rho_{1}(x,x')$ is visualized for $N=5$ bosons in a harmonic
trap ($h=0$, top row) and a double well of barrier strength $h=5$
(bottom). In the harmonic case, the system starts at $g=0$ with a
direct-product state $\psi=\phi_{0}^{\otimes N}$, i.e., with a density
matrix $\rho_{1}(x,x')=\phi_{0}(x)\phi_{0}^{*}(x')\propto e^{-R^{2}}e^{-r^{2}/4}$
in terms of $r=x-x'$ and $2R=x+x'$. From this point of view, the
system does not exhibit off-diagonal long-range order, which is simply
rooted in the fact that it is spatially bounded. Of course, it is
nonetheless in a coherent state and thus features \emph{weak} long-range
order in that $\rho_{1}(x,-x)\sim\sqrt{\rho(x)\rho(-x)}$ as $x\to\infty$.
This property persists so long as the correlations induced by the
interactions are weak enough for the system to remain in such a single-particle
state (the Gross-Pitaevskii regime), such as for $g=0.4$. For $g=4.7$,
however, the symmetry in $R$ and $r$ breaks up. The density profile
$\rho(x)\equiv\rho_{1}(x,x)$ flattens, and one can see that the off-diagonal
range is somewhat extended, too. However, as $g$ is increased further,
the support of $\rho_{1}(x,x')$ will concentrate more and more in
the central region $\{ x=x'\}$, where the typical fermionized profile
is recovered (cf. $g=194$). By contrast, the off-diagonal contributions
will be washed out, indicating the decoherence of the system. Still
it is noteworthy that even in this limit, a rest of coherence is preserved
in a faint checkerboard pattern. 

For the double well ($h=5$; bottom row), the situation is apparently
different. As always, the system exhibits coherence to begin with
(cf. $g=0.4$), only that the orbital is now delocalized in both minima
$\pm x_{0}$ and may be written as $\phi_{0}(x)=c[\varphi(x-x_{0})+\varphi(x+x_{0})]$.
Unlike the harmonic case, the off-diagonal range is not initially
increased but directly destroyed upon switching on $g$. While for
$g=4.7$, the density matrix $\rho_{1}(x,x')$ may still be thought
of as pertaining to two separate subsystems, it eventually reaches
the Tonks-Girardeau limit ($g=194$), where the only essential difference
toward $h=0$ consists in the density suppression at $x,x'=0$.

\subsection{Natural orbitals and their populations \label{sub:natorb}}

While, in principle, the \emph{full} density matrix $\rho_{1}(x,x')$
as studied in the previous section contains all the information about
fragmentation at the one-particle level, it is somewhat less amenable
to intuition. A handier criterion is offered by its spectral decomposition
(\ref{eq:rho1}) in terms of its \emph{natural orbitals} $\phi_{a}$
and their populations $n_{a}$, telling us how close the system is
to a pure one-orbital state.

\subsubsection{Natural populations as a measure of fragmentation}

\begin{figure}
\begin{center}\subfigure[]{\includegraphics[%
  width=6cm]{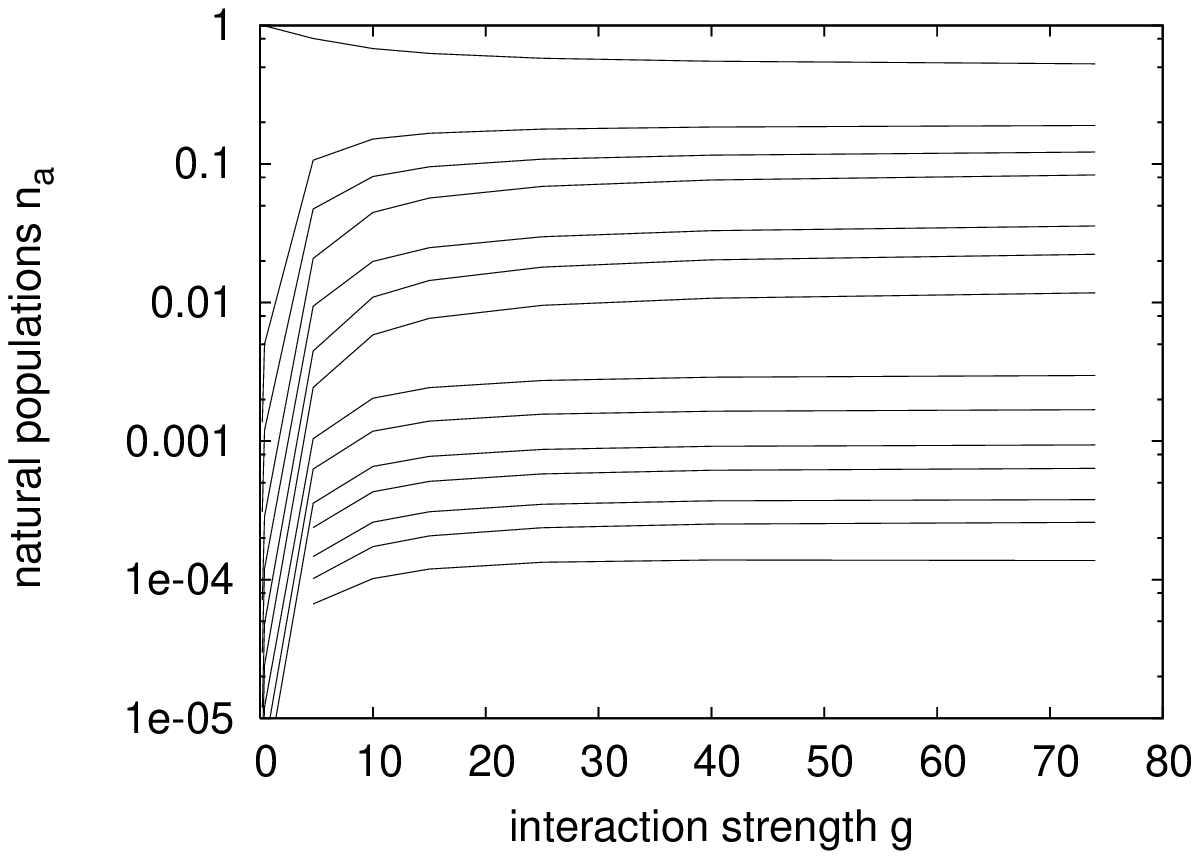}}\subfigure[]{\includegraphics[%
  width=6cm]{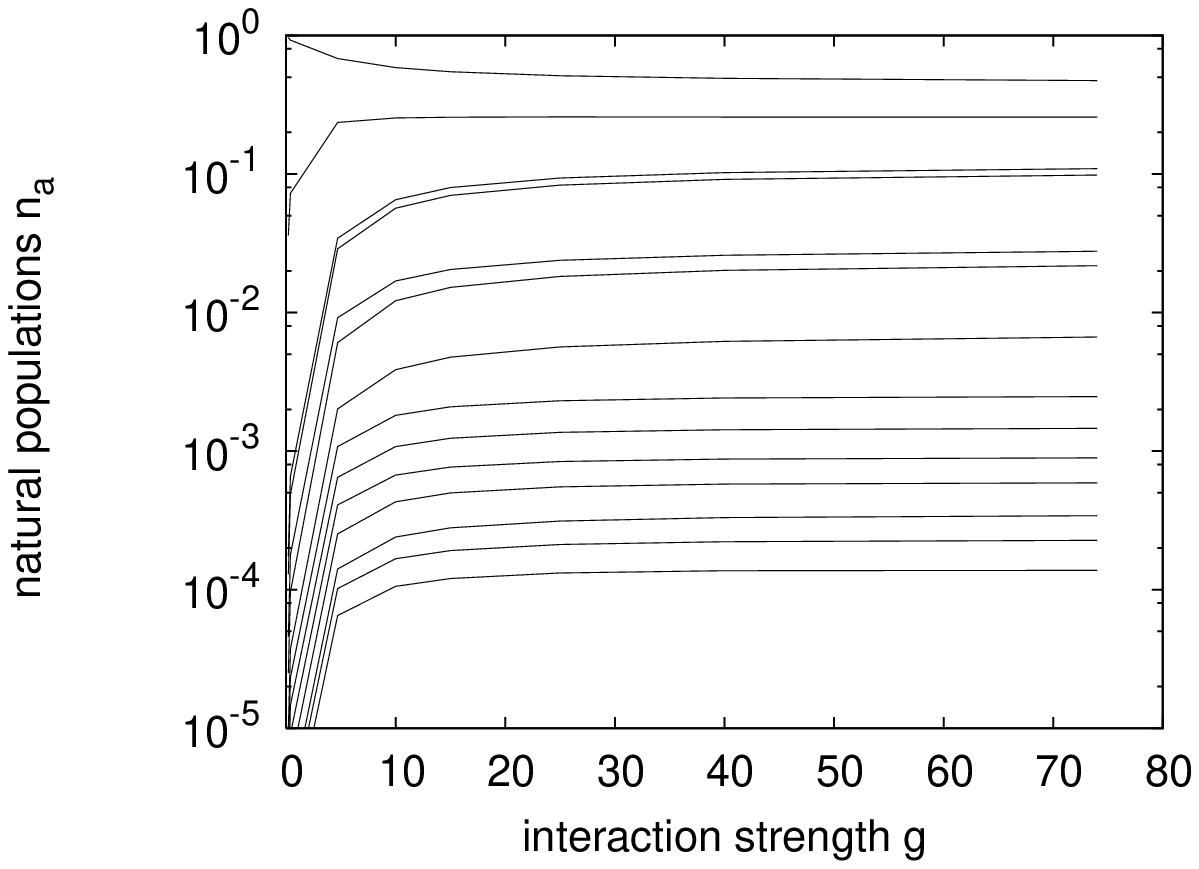}}\subfigure[]{\includegraphics[%
  width=6cm]{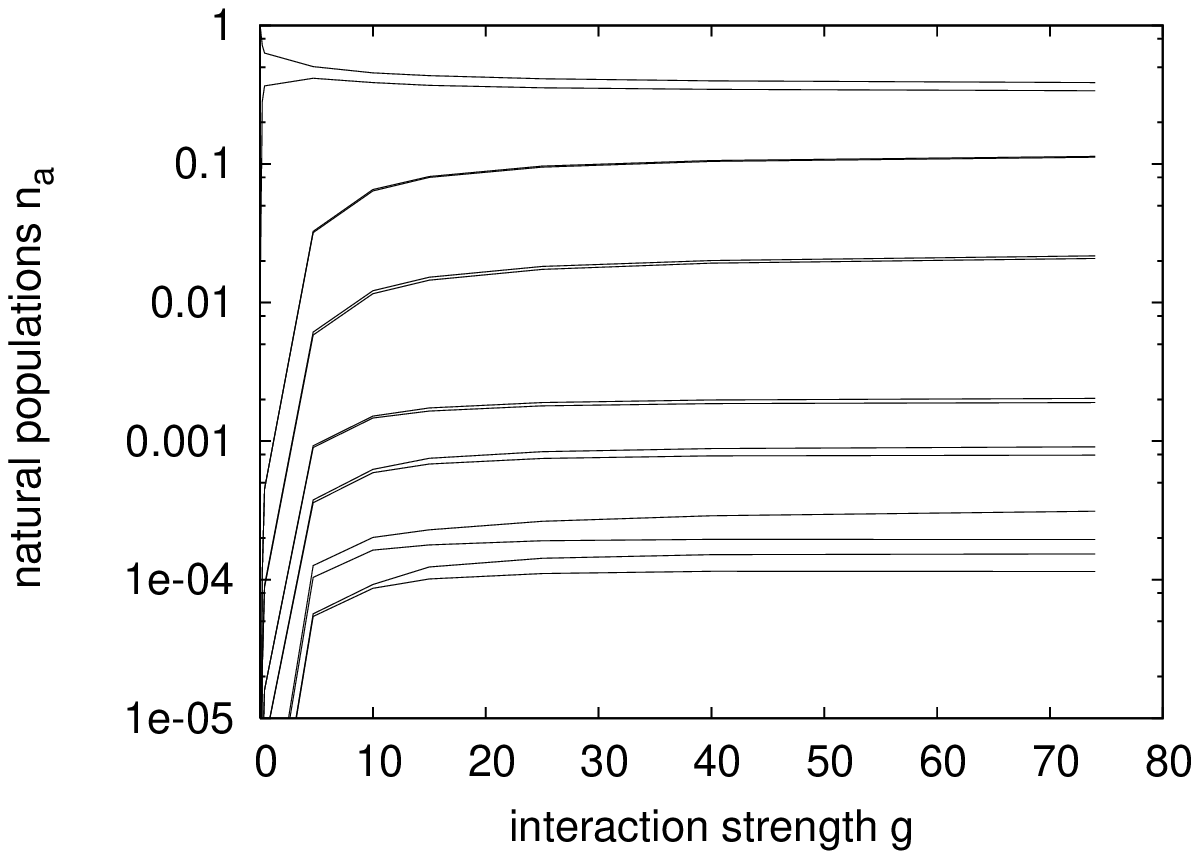}}\end{center}

\caption{Natural populations $n_{a}(g)$ $(a\le13)$ for $N=4$ bosons in
a harmonic trap (a) and in a double well with barrier height $h=5$
(b) , $h=10$ (c). \label{cap:natpop}}
\end{figure}
Figures~\ref{cap:natpop}(a-c) show typical plots of the natural
populations as the interaction is increased, $\{ n_{a}(g)\}$, for
four bosons and $h\in\left\{ 0,5,10\right\} $. Starting from $n_{0}=1$
for the non-interacting case, the lower lines rise steeply until they
end up saturating in a fermionized state at $g\to\infty$. Note that
this pattern is to some degree universal. In other words, it is roughly
detached from the specific shape of the trap, i.e, from what the underlying
\emph{orbitals} look like. This indicates why the set $\{ n_{a}\}$
lends itself as a handy criterion for fragmentation. The details of
the system are essentially encoded in (i) the exact sequence of $n_{a}$
in the Tonks-Girardeau limit, and (ii) in the transition between the
two extreme regimes $g=0$ and $g\to\infty$.

For the harmonic oscillator ($h=0$), the plot reveals a relatively
simple hierarchy. The value of $n_{0}$ decreases smoothly to its
Tonks-Girardeau limit $N^{-0.41}$ \cite{girardeau01}. All the remaining
populations increase dramatically up until $g\sim10$, and accumulate
in a more or less equidistant spacing (on a log scale). But even the
next-to-dominant weight $n_{1}$ is nowhere near the {}`condensate'
fraction $n_{0}$; the obvious gap between these two reflects the
difficulty to observe fragmentation in the harmonic oscillator as
compared to $h>0$. Note that the group of lines $\{ n_{0},\dots,n_{N-1}\}$
reveals a discernible separation from the lines below. This clearly
relates to the finding that the convergence of the energy $E(n)$
as a function of the number of orbitals, just to give an example,
gets strikingly better once $n\ge N$ \cite{zoellner06a,streltsov06}.
It is the accumulation of points $n_{a}(g)$ that makes for the utter
slowness of \emph{true} convergence pointed out in that work. 

For a barrier with height $h=5$, a little more structure can be identified
in the line sequence $n_{a}(g)$. The accumulation persists, but at
least the more populated orbitals $a$ seem to come in groups of two.
This will become clearer when looking into the natural orbitals. Even
more striking is the behavior of the second orbital's population,
$n_{1}$. It increases with $g$ much more rapidly than all others,
and it becomes comparable with $n_{0}$ already for modest $g\sim5$.
This scale separation between the pair $n_{0/1}$ and the rest is
in sharp contrast to the HO case. It gives a qualitative justification
of the two-mode approximation widely used in double-well systems.
To make these points even clearer, we have plotted the results for
a much higher barrier, $h=10$. Here $n_{1}$ {}`jumps' almost instantaneously
($g\ll1$) to a value of order $\frac{1}{2}$, whereas the remaining
occupations only catch up for $g\sim5$. It is in that regime that
the 2-mode model works brilliantly (see also Sec.~\ref{sub:approximations}).

The reason why fragmentation is facilitated when the central barrier
is raised is intuitively clear. The particles' tendency to separate
due to repulsion is usually obstructed by the higher costs of kinetic
and potential energy. The potential-energy barrier creates an additional
incentive for the bosons to separate. This has also been argued on
more quantitative grounds (see, e.g., \cite{spekkens99}). In a naive
single-particle picture, the energy gap $\Delta$ in a double well
between anti- and symmetric state, $\phi_{\pm}(x)=c[\varphi(x-x_{0})\pm\varphi(x+x_{0})]$,
vanishes as $h\to\infty$. It is thus far easier for the interaction
to bridge that gap for larger barriers, in particular compared to
the gap for $h=0$, $\Delta=1$.

Our final remark concerns the dependence on the atom number $N$.
For odd $N$, two features of the even-$N$ picture will differ. First,
the second mode is less relevant (e.g. for $N=5$, $n_{1}|_{g=4.7}=0.16$).
What is more, the separation between the first $N$ populations and
all others was found to be much smaller.  This backs up the intuitive
notion that, for odd $N$, fragmentation is seemingly impeded \cite{zoellner06a}.

\subsubsection{Natural orbitals\label{sub:Natural-orbitals}}

Even though the natural orbitals $(\phi_{a})$ are not of direct physical
importance, they are a valuable tool to gain some insight into the
process of fragmentation, as they determine both the spatial density
matrix $\rho_{1}(x,x')$ as well as the momentum density $\tilde{\rho}$,
to be discussed in the following subsection. In the uncorrelated case
$g=0$, the system is in a number state $|N,0,\dots\rangle$ and thus
the natural orbitals coincide with the single-particle eigenstates.
Since $V$ is a continuous perturbation, the orbitals $\phi_{a}$
will be somewhat distorted in the course of increasing $g$. For small
enough $g$---i.e., in the Gross-Pitaevskii regime---that modified
$\phi_{0}$ will suffice for an accurate description. Conversely,
if correlations are sufficiently influential, many orbitals will contribute
to $\rho_{1}$, and studying their interplay will illuminate our results
on the density matrix and the momentum distribution.

\paragraph*{Harmonic trap ($h=0$)}

\begin{figure}
\subfigure[]{\includegraphics[%
  width=7cm,
  keepaspectratio]{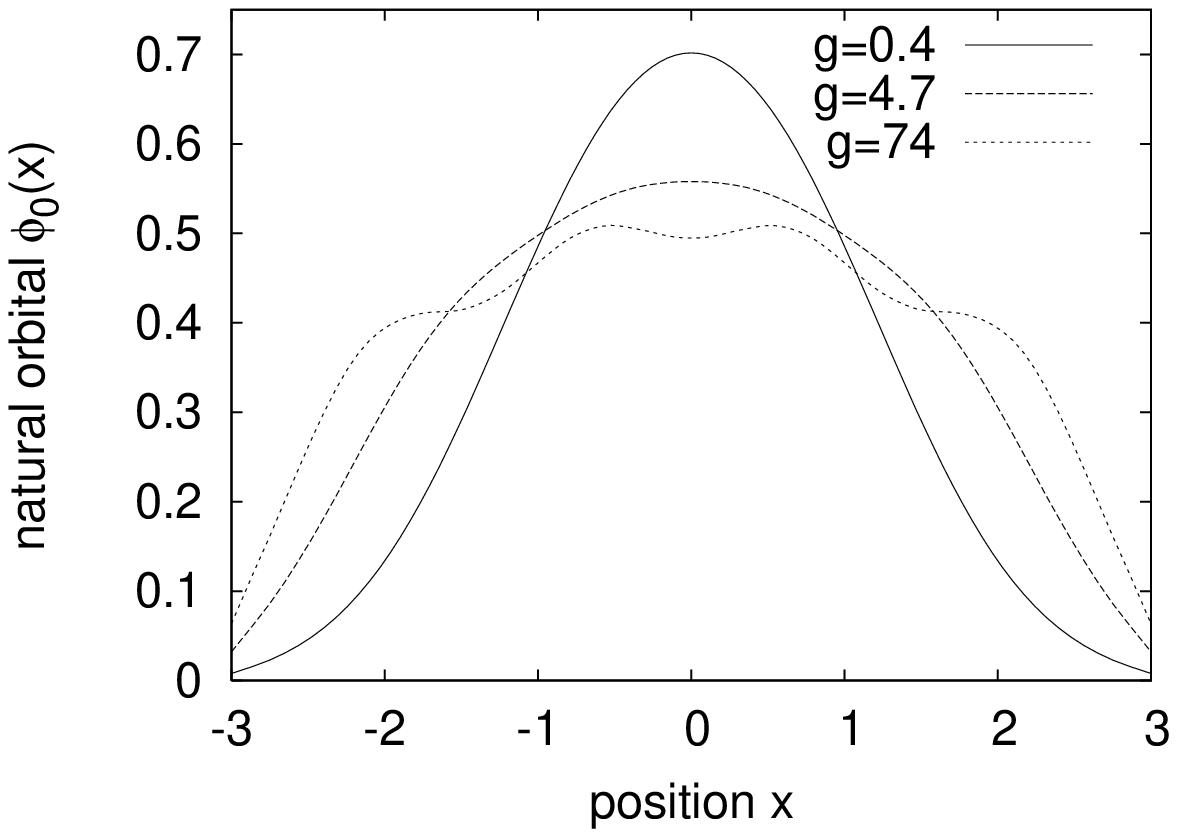}}\subfigure[]{\includegraphics[%
  width=7cm,
  keepaspectratio]{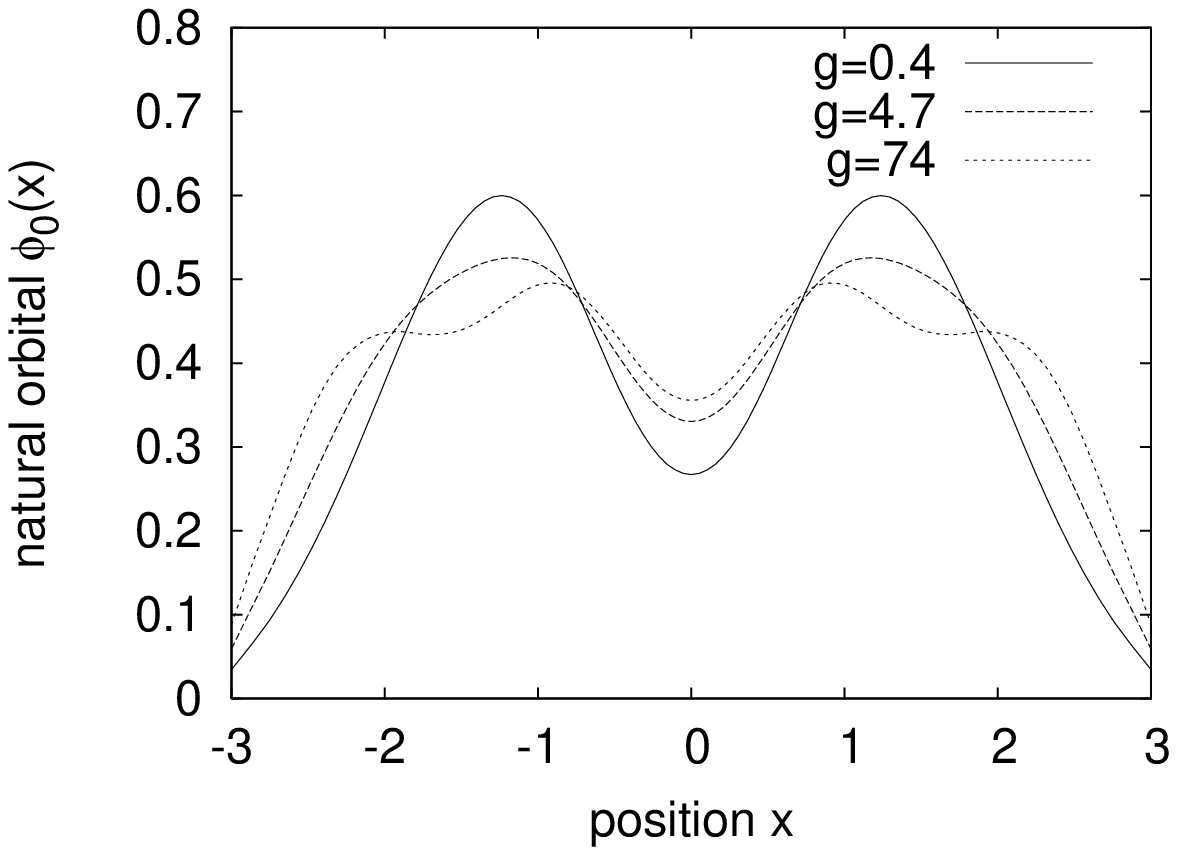}}

\subfigure[]{\includegraphics[%
  width=7cm,
  keepaspectratio]{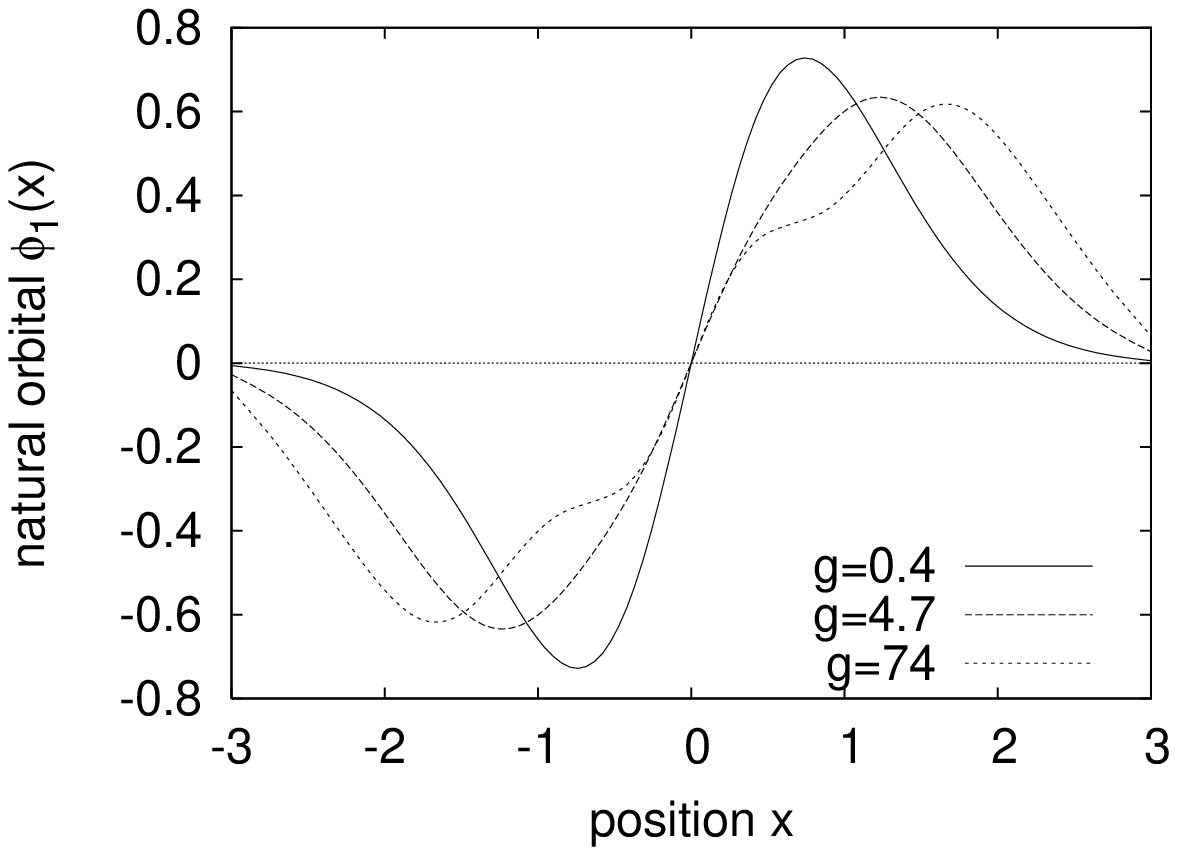}}\subfigure[]{\includegraphics[%
  width=7cm,
  keepaspectratio]{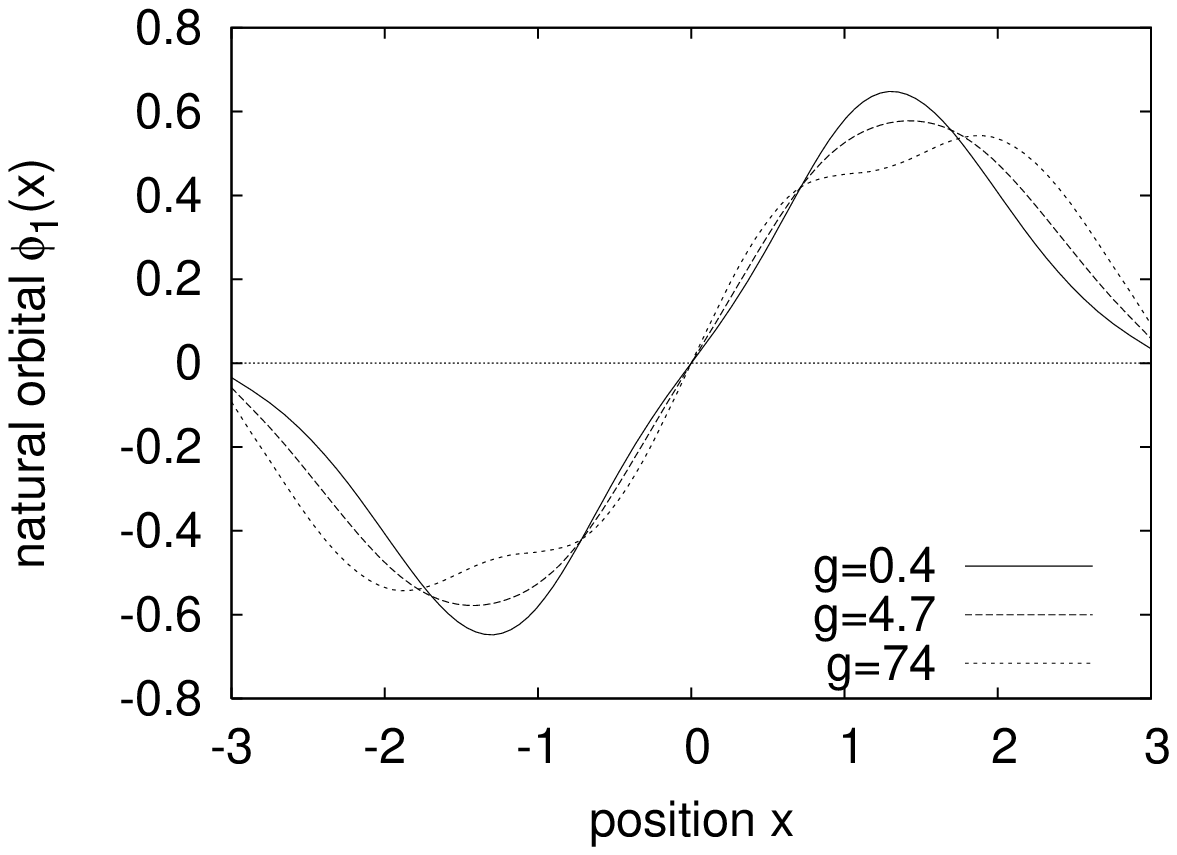}}

\caption{Natural orbitals $\phi_{a}$ for the case $N=4$. Top row: {}`condensate'
orbital $\phi_{0}$ for $h=0$ (a), $h=5$ (b). Bottom: $\phi_{1}$
for $h=0$ (c), $h=5$ (d). Each subfigure shows plots for the representative
interaction strengths $g=0.4$ (weak repulsion), $g=4.7$ (intermediate
regime), and $g=74$ (fermionization limit). \label{cap:natorb}}
\end{figure}
For the harmonic trap (Figs.~\ref{cap:natorb}a,c), the initial HO
function $\phi_{0}$ is only slightly flattened in the Gross-Pitaevskii
regime (cf. $g=0.4$). The onset of fragmentation not only smears
out the lowest orbital, but also admixes an antisymmetric HO-type
orbital $\phi_{1}$. In the fermionization limit, it is astonishing
that already $\phi_{0}$ exhibits all the features of the fermionized
density profile $\rho(x)$, that is, $N$ pronounced humps mirroring
the spatial isolation of the atoms. This is intelligible given that
$\phi_{0}$ still has a \emph{dominant} weight, which ought to be
contrasted with the philosophy of multi-orbital mean-field schemes
(see, e.g., Ref.~\cite{alon05}), where that pattern is produced
by $N$ spatially localized orbitals of \emph{equal} population. That
issue will be discussed in more detail in Sec.~\ref{sub:approximations}.

Interesting as the orbitals may be in their own right, they also prove
helpful in clarifying the decoherence found in Sec.~\ref{sub:ODLRO}.
The onset of fragmentation, as for $g=4.7$, leads to a broadened
diagonal profile $\rho_{1}(x,x)$, but not equally so for the off-diagonal
part. That is simply because the $\phi_{a}$ have alternate parity
$\left(-1\right)^{a}$, and thus the admixture of another orbital
leads to $\rho_{1}(x,-x)=\sum_{a}\left(-1\right)^{a}n_{a}\left|\phi_{a}(x)\right|^{2}$.
Hence the fragmentation into different orbitals tends to deplete the
off-diagonal as compared to the diagonal density. For $g=4.7$, this
effect is still tiny as $n_{1}\sim0.1$ only, and therefore outweighed
by the altogether extended support of $\phi_{0}$. However, as more
and more orbitals are mixed, as is the case in the fermionization
limit (see $g=74$), this \emph{decoherence} attains its full impact.
We remark that the faint checkerboard pattern (Fig.~\ref{cap:ODLRO})
is still rooted in the dominance of the lowest orbital, $n_{0}\simeq N^{-0.41}$.

\paragraph*{Double well ($h=5$)}

In the case of a central barrier (Figs.~\ref{cap:natorb}b,d), the
natural orbitals in the non-interacting limit will again be the single-particle
eigenstates, approximately the symmetric and anti-symmetric states
$\phi_{\pm}$. In the Gross-Pitaevskii regime ($g=0.4$), the lowest
orbital is only marginally flattened due to interactions, but a tiny
reduction of the off-diagonal peaks $\rho_{1}(\pm x_{0},\mp x_{0})$
hints already at a minor admixture of the antisymmetric $\phi_{1}$.
For $g=4.7$, fragmentation has set in, not only smearing out the
orbitals $\phi_{0/1}$ ---and thus the diagonal profile---but along
the way washing out the off-diagonal long-range order almost completely.
As emphasized before, the fermionization pattern tends to be generic
for different $h$, which reflects both in the density matrix as well
as in the natural orbitals.

\subsection{Momentum density\label{sub:Momentum-density}}

The discussion so far focused on rather abstract aspects of the one-body
correlations. Yet it can help us cast a light on an experimentally
more amenable quantity, the momentum density.

\paragraph*{Harmonic trap ($h=0$)}

\begin{figure}
\subfigure[]{\includegraphics[%
  width=7.5cm,
  keepaspectratio]{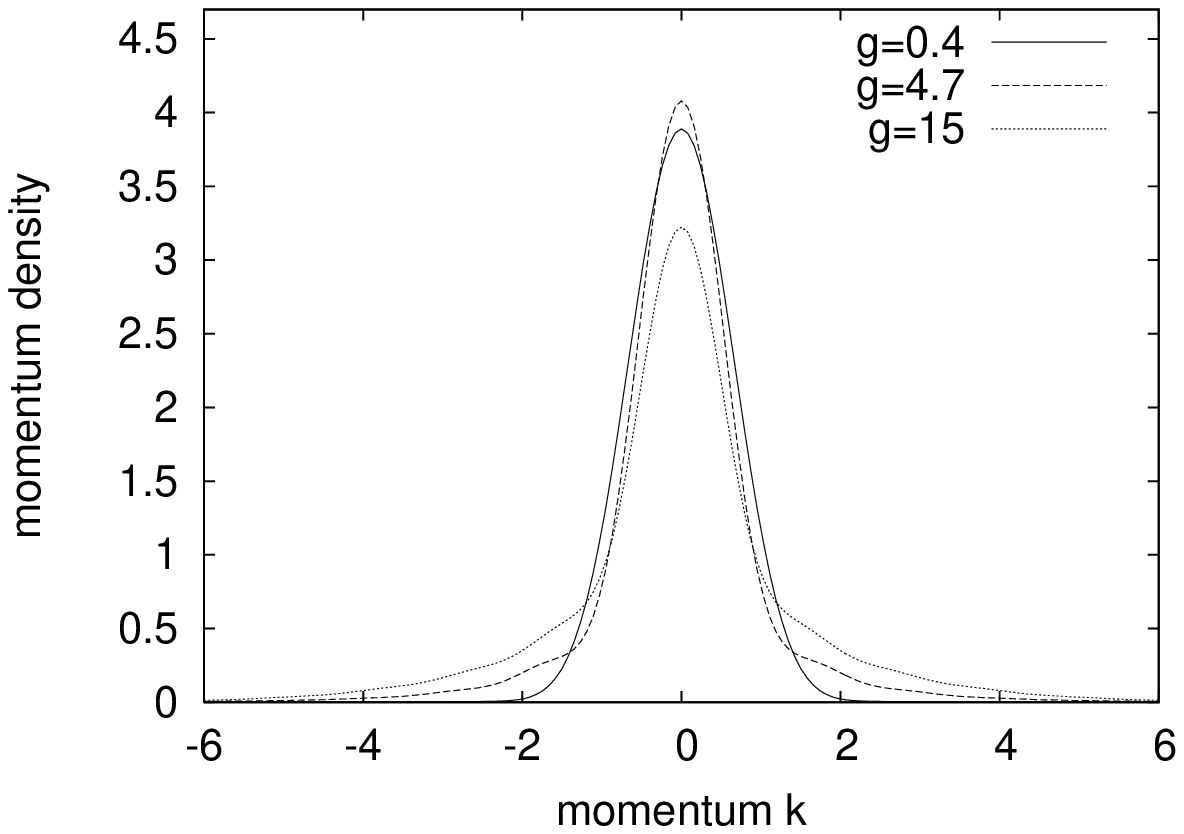}}\subfigure[]{\includegraphics[%
  width=7.5cm,
  keepaspectratio]{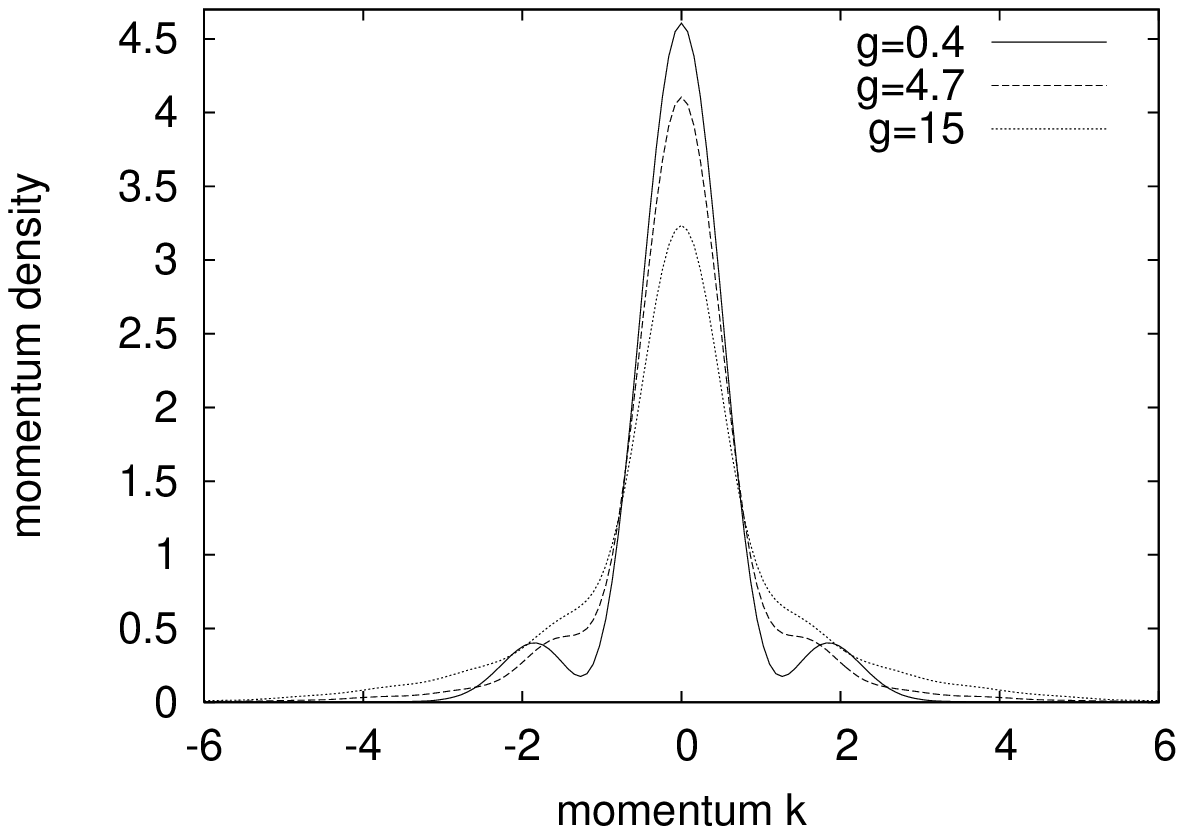}}

\subfigure[]{\includegraphics[%
  width=7.5cm,
  keepaspectratio]{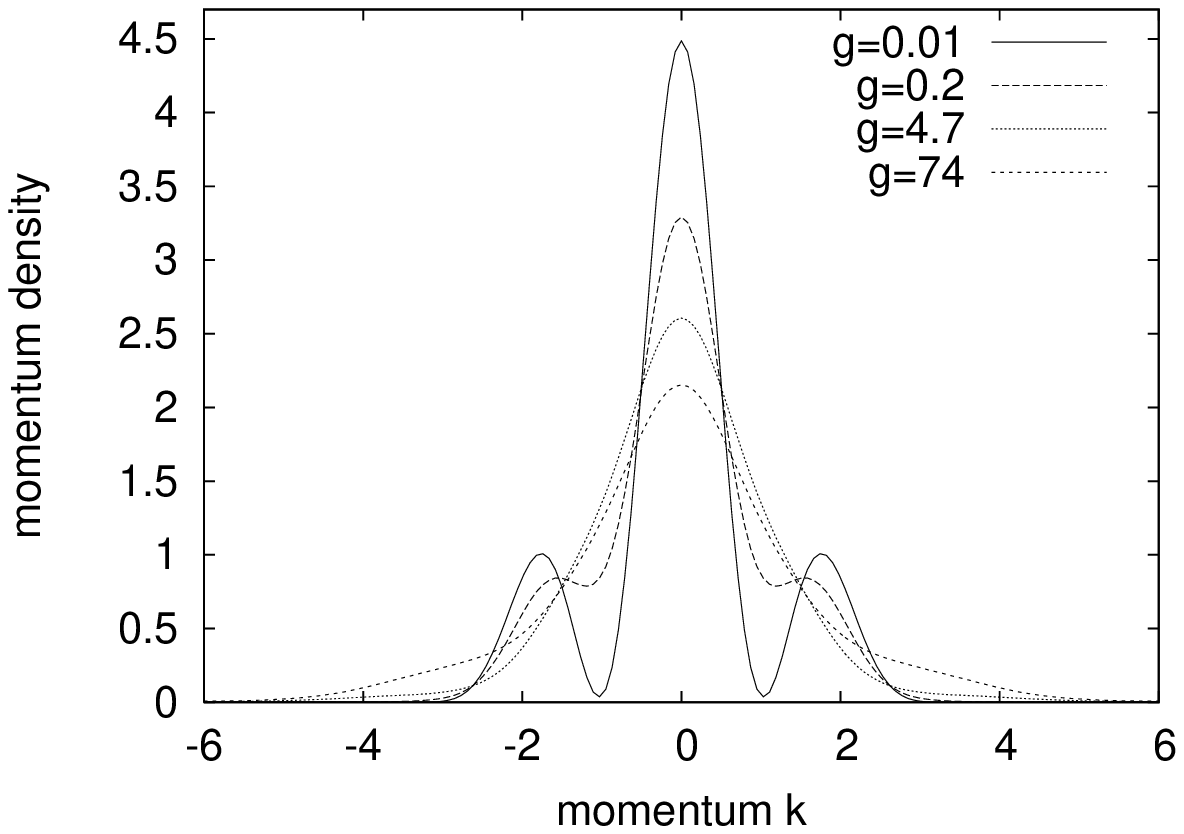}}

\caption{Momentum density $\tilde{\rho}(k)$ for (a-b) $N=5$ bosons in a
double-well trap of barrier height $h$. (left: $h=0$, right: $h=5$);
shown are the interaction strengths $g=0.4,\,4.7,\,15$. (c) The case
$N=4,h=10$ for $g=0.01,\,0.2,\,4.7,\,74$. \label{cap:DW_k}}
\end{figure}
For this case, the momentum distribution has recently been computed
(\cite{deuretzbacher06}; see also Ref.~\cite{astrakharchik03}),
yet we plot it in Figure~\ref{cap:DW_k} for comparison. It evolves
from a Gaussian $\tilde{\rho}(k)/2\pi=\pi^{-1/2}e^{-k^{2}}$ at $g=0$
(with a maximum at $\tilde{\rho}(0)=.35\dots$) to a slightly sharper
peak, here depicted for $g=.4$. This squares with the broadened natural
orbital $\phi_{0}$ in that regime, as found in Sec.~\ref{sub:Natural-orbitals}.
In the same light one sees that for $g=4.7$, where fragmentation
has set in, the peak at $k=0$ is even more pronounced, while $\tilde{\rho}(k)$
has also developed a long-range tail. Both observations are easily
accounted for. The $k=0$ behavior, for one thing, was argued to correspond
to the off-diagonal long-range behavior of $\rho_{1}(x,x')$ in Sec.~\ref{sub:Fragmentation}.
This fits in with our observation that the off-diagonal range was
indeed extended in that $g$-regime, as seen in Fig.~\ref{cap:ODLRO}. 

The asymptotics $k\to\infty$ is in turn determined by the short-range
interaction, which is known to culminate in the $k^{-4}$ tail for
$g\to\infty$. This latter consequence is in fact confirmed here (see
$g=15$). Moreover, notice that the $k=0$ peak is bound to diminish.
In other words, the momentum spectrum is redistributed toward higher
$k$, in accordance with the reduction of off-diagonal long-range
order. This fact stands in marked contrast to the homogeneous system,
which in the Tonks-Girardeau limit had an infrared divergence $\tilde{\rho}(k)=O(k^{-1/2})$.
The seeming contradiction is owed to the fact that we deal with a
bounded system, which cannot display true long-range order.

\paragraph*{Double well ($h=5$)}

The momentum spectrum for a double well looks quite different from
the start ($g=0.4$): it exhibits two sidelobes. This can be explained
by the symmetric orbital $\phi_{0}(x)=c[\varphi(x-x_{0})+\varphi(x+x_{0})]$,
which leads to a cosine-type modulation of $\tilde{\rho}$ due to
$\hat{\phi}_{0}(k)=c\cos(kx_{0})\hat{\varphi}(k)$. These sidelobes
get even more distinct as $h\to\infty$ (see Fig.~\ref{cap:DW_k}c).
With increasing repulsion ($g=4.7$), there are two competing effects.
On the one hand, the orbitals are flattened a little, which should
result in a slightly sharper momentum distribution. It turns out,
though, that the effect of fragmentation outperforms the former one
even for tiny interactions: admixing an anti-symmetric orbital $\phi_{1}$
adds a $\sin(kx_{0})$-type modulation, thus washing out the sidelobes
as well as the central peak. Note that this effect is even more striking
for $h=10$, where it kicks in already for $g=0.2$. In other words,
the signature of the Gross-Pitaevskii regime in the harmonic trap---the
initial sharpening of the $k=0$ peak---is lost in the case of a sufficiently
pronounced double well.

Along the lines of the remarks in the previous paragraph, we mention
that the behavior for large correlations $g$ is again universal as
far as the $k^{-4}$ tail for $k\to\infty$ is concerned. It also
has a reduced peak for zero momentum, in accordance with the loss
of long-range order found in Sec.~\ref{sub:ODLRO}.

\section{Two-body correlations and discussion \label{sec:2p-correlation}}

The focal point of our discussion so far has been the one-particle
density matrix, and related quantities. However useful they are in
studying fragmentation, they are by construction ignorant of an essential
ingredient: the two-body correlations, which have of course been traced
out in the definition of $\rho_{1}$. Studying the (diagonal) two-particle
density $\rho_{2}(x_{1},x_{2})$ may thus promise to yield an intriguing
look behind the scenes of the one-particle picture. We will round
off this section by commenting on the relation of our results to approximate
methods such as multi-orbital mean-field theory and the two-mode model.

\subsection{Two-particle correlations}

\begin{figure}
\includegraphics[%
  width=5cm,
  keepaspectratio]{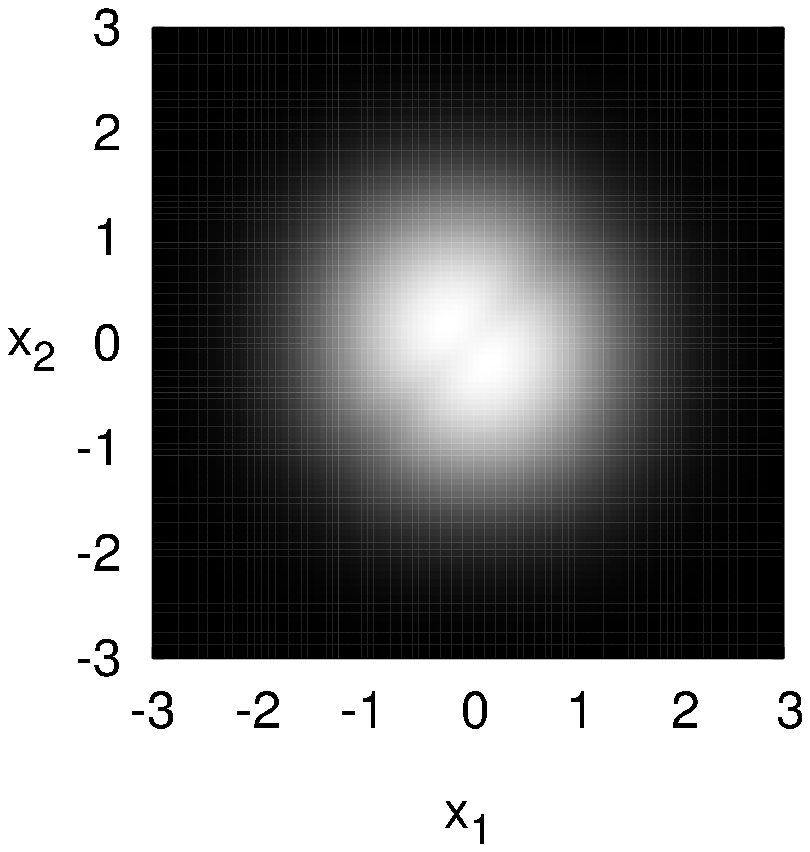}\includegraphics[%
  width=5cm,
  keepaspectratio]{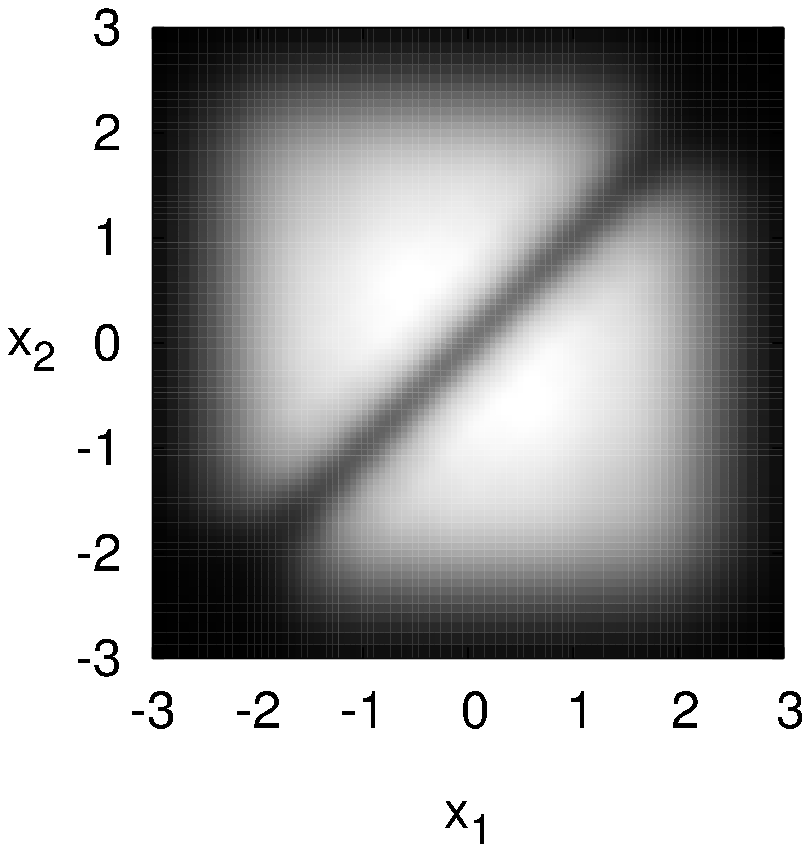}\includegraphics[%
  width=5cm,
  keepaspectratio]{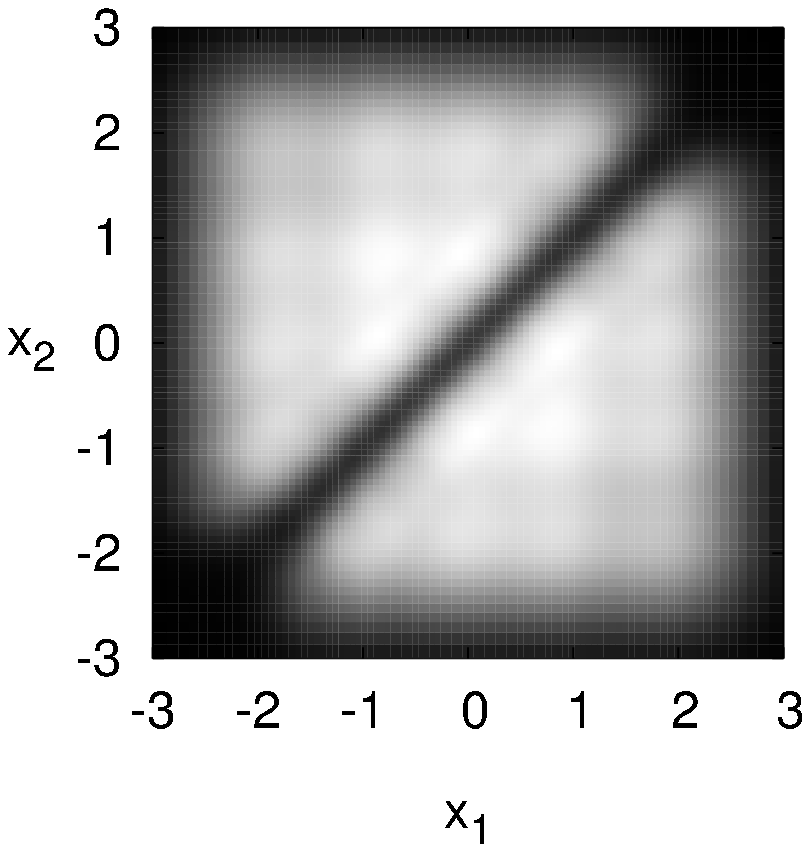}

\includegraphics[%
  width=5cm,
  keepaspectratio]{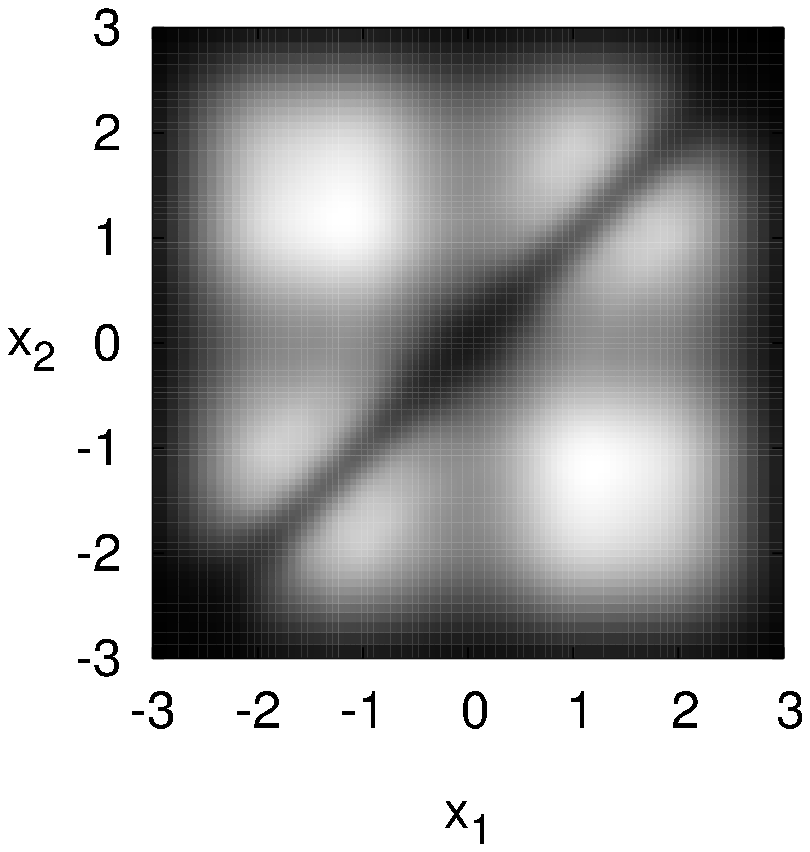}\includegraphics[%
  width=5cm,
  keepaspectratio]{figures/p5d1_DW.2.h5.eps}\includegraphics[%
  width=5cm,
  keepaspectratio]{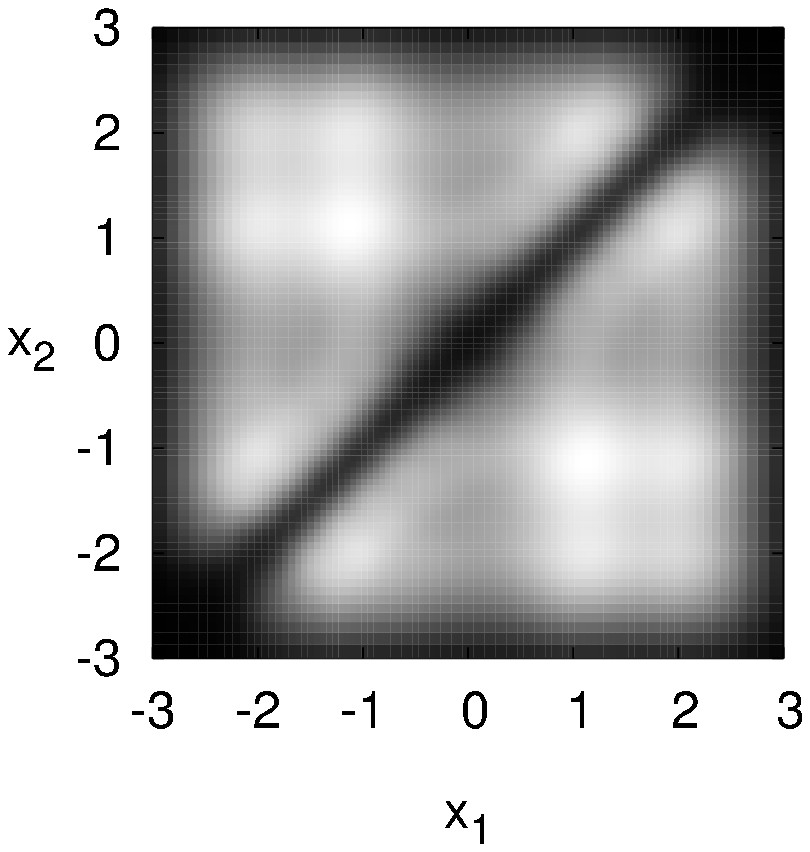}

\caption{2-particle density $\rho_{2}(x_{1},x_{2})$ for $N=5$ bosons in
a double-well trap of barrier height $h$. Top row: $h=0$, bottom
row: $h=5$; shown are the interaction strengths $g=0.4,\,4.7,\,10$
from left to right. \label{cap:density2}}
\end{figure}
For weak enough interactions, $g\to0$, the system is in an uncorrelated
state characterized by $\rho_{2}(x_{1},x_{2})=\rho(x_{1})\rho(x_{2})$.
The first effect of the two-body interaction is to distort the one-particle
density governed by the Gross-Pitaevskii eq., while from some point
on $\rho_{2}$ will reflect the correlations that are introduced by
$V(x_{i}-x_{j})$. As the pair-interaction energy is $\mathrm{tr}(V\rho_{2})\stackrel{\sigma\to0}{\sim}g\int dx\rho_{2}(x,x)$,
keeping it low amounts to depleting the {}`correlation diagonal'
$\{ x_{1}=x_{2}\}$. 

For $g=0.4$ (Fig.~\ref{cap:density2}), a closer look reveals exactly
that. In the case $h=0$, a slight dip along the diagonal is formed
as the density is being smeared out a little. In turn, for the double
well ($h=5$), the density is pumped into the off-diagonal peaks,
which is indicative of a correlated state. Note that this {}`dip'
on the correlation diagonal is a feature of the two-body picture;
in the one-particle density $\rho=\int dx_{2}\rho_{2}(\cdot,x_{2})$
it is smoothed out and thus much less visible. As $g$ is increased
($g=4.7$), we are in the regime of fragmentation. Here the wave function
develops marked minima at points of collision, $x_{i}=x_{j}$, as
exemplified for two atoms in a harmonic trap \cite{cirone01}. This
carries over to $\rho_{2}$, where a characteristic correlation hole
emerges, cutting the plot into halves. Despite that, the overall pattern
still bares a resemblance to the non-interacting case. This changes
when the system approaches fermionization, cf. $g=10$. Here we encounter
the simultaneous splitting into humps familiar from the one-particle
density as presented in Ref.~\cite{zoellner06a}. In this context,
these wiggles signify that, if one boson resides at $x_{1}$, then
any second one is likely to be found at $N-1$ distinct spots $x_{2}$,
just not at $x_{2}=x_{1}$. The exact distribution depends on the
trap, of course. For $h=0$, the checkerboard pattern is quite regular
and has a larger amplitude about $x=0$, while for $h=5$ the peaks
are in a way packed into either well but suppressed in the center.

\subsection{Relation to approximation schemes\label{sub:approximations}}

Let us stop to wrap up what we have found and work out the key points
by contrasting them with two well-known approximations. The transition
from a macroscopic state $\phi_{0}^{\otimes N}$ to a fermionized
state follows different pathways for different traps. For a harmonic
trap, the lowest orbital retains its singular importance for any interaction
strength. In a double well, as the barrier height tends to infinity,
the route passes through a configuration fairly well approximated
by two single-particle states $\phi_{a}$, i.e., a number state $|N/2,N/2\rangle$
(or generally a superposition of states $|n'_{0},n'_{1}\rangle$).
For higher $g$, in turn, the system will be fermionized, and the
first $N$ single-particle orbitals will have dominant weights $n_{a}$.

The stopover near the fragmented state $|N/2,N/2\rangle$ in the double
well is the essence of the commonplace \emph{two-mode approximation}
(see, e.g., \cite{spekkens99}). It can be recovered in MCTDH by restricting
the number of single-particle functions to $n=2$ (see \ref{sec:method}),
which sets the subspace to $\mathrm{span}\{|n'_{0},n'_{1}\rangle\mid\sum_{a}n_{a}^{\prime}=N\}$.
Needless to say, it becomes only exact in the limit $h\to\infty$
and $g\to0^{+}$. Never is it able to describe anything but that coarse-grained
fragmentation into two simple fragments, let alone the typical fermionization
pattern in $\rho(x)$, the correlation hole evidenced in $\rho_{2}(x_{1},x_{2})$,
or the short-range-driven $k^{-4}$ tail evidenced in the momentum
spectrum.

\begin{figure}
\begin{center}\includegraphics[%
  width=7cm,
  keepaspectratio]{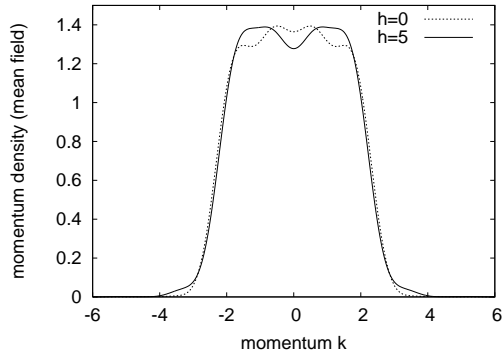}\end{center}

\caption{The spurious momentum density $\tilde{\rho}(k)$ as obtained within
the restriction to $n=N=4$ orbitals in the Tonks-Girardeau limit.
Plotted are the results for the harmonic trap ($h=0$) and the double
well $(h=5$) for large interactions $g$. The states correspond to
number states $|1_{0},\dots,1_{N-1}\rangle$. \label{cap:MF_k}}
\end{figure}

The simple two-mode state above, $|n'_{0},n'_{1}\rangle$, is contained
in a general \emph{multi-orbital mean-field} theory, which yields
the variationally best number state \cite{alon05,cederbaum03}. Among
its most impressive successes has been a \emph{}description of the
evolution from the Gross-Pitaevskii regime to fermionization. The
latter one was modeled by a state with $n_{a}'=1$ $(a<N)$. We have
pointed out \cite{zoellner06a} that this state emerges from our approach
as a robust yet spurious solution if the Hilbert space is restricted
to $\mathrm{span}\{\Phi_{J}\mid j_{\kappa}\le N\}$. Also, we have
argued that the number state gets all local quantities (the reduced
densities as well as the energy) \emph{about} right. By contrast,
it is not designed to reproduce correlation-sensitive functions, such
as the off-diagonal density matrix. Even though the use of several
orbitals also serves to destroy off-diagonal long-range order (which
hinges on the superposition of different orbitals, as delineated in
\ref{sub:Fragmentation}), the incorrect populations $(n_{a})$ make
for a \emph{fermionic} rather than a \emph{bosonic} momentum density,
as displayed in Fig.~\ref{cap:MF_k}. It resembles the spatial distribution
and not the $k=0$-peaked and long-range structure in Fig.~\ref{cap:DW_k}.
This reflects the fact that the mean-field approach cannot possibly
recover the correct short-range behavior, which would require explicit
correlations or, equivalently, the superposition of \emph{several}
single-particle configurations $\Phi_{J}$. Instead it only mimics
the spatial separation. 

Lastly, one should be aware that for a state with $n_{a}=1\,\forall a$,
the spectrum of $\rho_{1}$ is entirely degenerate; hence the eigenvectors
$\phi_{a}$ are only defined up to unitary transform $(U_{ab})$.
In this light they might as well be thought of as spatially localized,
as opposed to distorted oscillator functions ($h=0$) or anti-/symmetric
orbitals ($h=5$).%

\section{Conclusions and outlook}

The focus of this work have been correlation aspects of the numerically
exact ground state of few atoms in different double-well traps. This
way we have extended the more intuitive notion of the pathway from
the weakly interacting regime via fragmentation to fermionization
as described in our previous work \cite{zoellner06a}. On the other
hand, this article closes the gap between the non-interacting and
the Tonks-Girardeau limit, for which aspects such as off-diagonal
long-range order and the momentum spectrum have been understood, and
shows how these very different cases connect. Our method is based
on the Multi-Configuration Time-Dependent Hartree code, whose efficient
variational approach allows us to compute the ground state to a high
accuracy.

As one key result, we have highlighted the relation of the fragmentation
process to the diminishing of off-diagonal long-range order in the
one-body density matrix. That mechanism has been explained in terms
of the eigenvectors of the density matrix, the natural orbitals. These
also allow us to relate the loss of long-range order to the momentum
spectrum, whose markedly peaked structure in the non-interacting case
is stretched into a characteristic high-momentum tail for stronger
interactions. Moreover, we show how the populations of all natural
orbitals evolve, which not only illuminates how the fragmentation
mechanism is altered as the trap is turned from a harmonic one to
a double well, but also casts a light on the validity of the two-mode
model. Finally, we have laid out in more detail the two-body nature
of the correlations, which reflects in the formation of a {}`correlation
hole' and culminates in the onset of the familiar checkerboard pattern.
This goes well beyond the scope of approximation schemes inspired
by a single-particle picture.

With these investigations, the analysis of the ground state of few-boson
systems in double-well traps may be considered complete. Future extensions
of these studies appear obvious. For one thing, the addition of more
wells to the trap is of interest. This touches on the question of
the few-body analog of an optical lattice, and its related effects
such as the superfluid/Mott-insulator transition. On the other hand,
a physically thrilling situation would involve not only the ground
state, but also excitations, and eventually as well looking into the
dynamics of the system. Given the richness of these fields on the
many-atom level, the detailed study of few atoms promises a wide range
of applications. All these efforts may serve as a bridge toward a
better control of ultracold few-body systems.

\begin{acknowledgments}
Financial support from the Landesstiftung Baden-Württemberg in the
framework of the project {}`Mesoscopics and atom optics of small
ensembles of ultracold atoms' is gratefully acknowledged by PS and
SZ. The authors also appreciate A.~Streltsov's helpful comments and
thank O.~Alon for illuminating discussions.

\end{acknowledgments}
\bibliographystyle{prsty}
\bibliography{/home/sascha/paper/pra/DW/phd,/home/sascha/bib/mctdh}

\begin{thebibliography}{10}

\bibitem{pitaevskii}
L. Pitaevskii and S. Stringari, {\em Bose-Einstein Condensation} (Oxford
  University Press, Oxford, 2003).

\bibitem{dalfovo99}
F. Dalfovo, S. Giorgini, L. Pitaevskii, and S. Stringari, Rev. Mod. Phys. {\bf
  71},  463  (1999).

\bibitem{pethick}
C.~J. Pethick and H. Smith, {\em Bose-Einstein condensation in dilute gases}
  (Cambridge University Press, Cambridge, 2001).

\bibitem{leggett01}
A.~J. Leggett, Rev. Mod. Phys. {\bf 73},  307  (2001).

\bibitem{Olshanii1998a}
M. Olshanii, Phys. Rev. Lett. {\bf 81},  938  (1998).

\bibitem{girardeau60}
M. Girardeau, J. Math. Phys. {\bf 1},  516  (1960).

\bibitem{vaidya79}
H.~G. Vaidya and C.~A. Tracy, Phys. Rev. Lett. {\bf 42},  3  (1979).

\bibitem{minguzzi02}
A. Minguzzi, P. Vignolo, and M.~P. Tosi, Phys. Lett. A {\bf 294},  222  (2002).

\bibitem{kinoshita04}
T. Kinoshita, T. Wenger, and D.~S. Weiss, Science {\bf 305},  1125  (2004).

\bibitem{paredes04}
B. Paredes {\it et~al.}, Nature {\bf 429},  277  (2004).

\bibitem{girardeau01}
M. Girardeau, E.~M. Wright, and J.~M. Triscari, Phys. Rev. A {\bf 63},  033601
  (2001).

\bibitem{papenbrock03}
T. Papenbrock, Phys. Rev. A {\bf 67},  041601  (2003).

\bibitem{cirone01}
M.~A. Cirone, K. G\'{o}ral, K. Rzazewski, and M. Wilkens, J. Phys. B {\bf 34},
  4571  (2001).

\bibitem{hao06}
Y. Hao, Y. Zhang, J.~Q. Liang, and S. Chen, Phys. Rev. A {\bf 73},  063617
  (2006).

\bibitem{sakmann05}
K. Sakmann, A.~I. Streltsov, O.~E. Alon, and L.~S. Cederbaum, Phys. Rev. A {\bf
  72},  033613  (2005).

\bibitem{alon05}
O.~E. Alon and L.~S. Cederbaum, Phys. Rev. Lett. {\bf 95},  140402  (2005).

\bibitem{masiello05}
D. Masiello, S.~B. McKagan, and W.~P. Reinhardt, Phys. Rev. A {\bf 72},  063624
   (2005).

\bibitem{streltsov06}
A.~I. Streltsov, O.~E. Alon, and L.~S. Cederbaum, Phys. Rev. A {\bf 73},
  063626  (2006).

\bibitem{deuretzbacher06}
F. Deuretzbacher, K. Bongs, K. Sengstock, and D. Pfannkuche, cond-mat/0604673
  (2006).

\bibitem{zoellner06a}
S. Z{\"o}llner, H.-D. Meyer, and P. Schmelcher, quant-ph/0605210  (2006).

\bibitem{mey03:251}
H.-D. Meyer and G.~A. Worth, Theor.\ Chem.\ Acc. {\bf 109},  251  (2003).

\bibitem{mey98:3011}
H.-D. Meyer,  in {\em {T}he {E}ncyclopedia of {C}omputational {C}hemistry},
  edited by P. v.~R.~Schleyer {\it et~al.} (John Wiley and Sons, Chichester,
  1998), Vol.~5, pp.\ 3011--3018.

\bibitem{bec00:1}
M.~H. Beck, A. J{\"a}ckle, G.~A. Worth, and H.-D. Meyer, Phys.\ Rep. {\bf 324},
   1  (2000).

\bibitem{spekkens99}
R.~W. Spekkens and J.~E. Sipe, Phys. Rev. A {\bf 59},  3868  (1999).

\bibitem{Penrose56}
O. Penrose and L. Onsager, Phys. Rev. {\bf 104},  576  (1956).

\bibitem{lieb03}
E.~H. Lieb, R. Seiringer, and J. Yngvason, Phys. Rev. Lett. {\bf 91},  150401
  (2003).

\bibitem{yang62}
C.~N. Yang, Rev. Mod. Phys. {\bf 34},  694  (1962).

\bibitem{yukalov05}
V.~I. Yukalov and M.~D. Girardeau, cond-mat/0507409  (2005).

\bibitem{mey90:73}
H.-D. Meyer, U. Manthe, and L.~S. Cederbaum, Chem.\ Phys.\ Lett. {\bf 165},  73
   (1990).

\bibitem{mctdh:package}
G.~A. Worth, M.~H. Beck, A. J{\"a}ckle, and H.-D. Meyer, The {MCTDH} {P}ackage,
  {V}ersion 8.2, (2000). H.-D. Meyer, {V}ersion 8.3 (2002). {S}ee
  http://www.pci.uni-heidelberg.de/tc/usr/mctdh/.

\bibitem{kos86:223}
R. Kosloff and H. Tal-Ezer, Chem.\ Phys.\ Lett. {\bf 127},  223  (1986).

\bibitem{meyer06}
H.-D. Meyer, F.~L. Qu\'{e}r\'{e}, C. L\'{e}onard, and F. Gatti, Chem. Phys.  in
  press  (2006).

\bibitem{lig85:1400}
J.~C. Light, I.~P. Hamilton, and J.~V. Lill, J.~Chem.\ Phys. {\bf 82},  1400
  (1985).

\bibitem{astrakharchik03}
G.~E. Astrakharchik and S. Giorgini, Phys. Rev. A {\bf 68},  031602  (2003).

\bibitem{cederbaum03}
L.~S. Cederbaum and A.~I. Streltsov, Phys. Lett. A {\bf 318},  564  (2003).

\end{thebibliography}

\end{document}